\documentclass[twocolumn,floats,eqsecnum,pra]{revtex4-1}

\usepackage{amsmath,amssymb}

\def \Zs {\mathbf{Z}}

\def \calS {\mathcal{S}}

\def \te  {\widetilde{e}_2}
\def \tee {\widetilde{e}_3}

\def \np {N_\phi}

\newcommand {\be}{\begin{equation}}
\newcommand {\ee} {\end{equation}}
\newcommand {\bea}{\begin{eqnarray}}
\newcommand {\eea} {\end{eqnarray}}
\newcommand{\non}{\nonumber}

\newcommand{\qdegen}[2]{\genfrac{\{}{\}}{0pt}{}{#1}{#2}}

\begin{document}


\title{Entanglement subspaces, trial wavefunctions, and special Hamiltonians\\ in the fractional quantum Hall effect}
\author{T.S. Jackson$^1$, N. Read$^2$, and S.H. Simon$^3$}
\affiliation{$^1$Department of Mathematics, University of Pennsylvania, Philadelphia, PA 19104, USA\\
$^2$Department of Physics, Yale
University, P.O. Box 208120, New Haven, CT 06520-8120, USA\\
$^3$Rudolf Peierls Centre for Theoretical Physics, University of Oxford, Oxford OX1 3NP, UK}
\date{August 13, 2013}

\begin{abstract}
We consider spaces of trial wavefunctions for ground states and edge excitations in the fractional
quantum Hall effect that can be obtained
in various ways. In one way, functions are obtained by analyzing the entanglement of the ground
state wavefunction, partitioned into two parts. In another, functions are defined by the way in which
they vanish as several coordinates approach the same value, or by a projection-operator Hamiltonian
that enforces those conditions. In a third way, functions are given by conformal blocks from a
conformal field theory (CFT). These different spaces of functions are closely related. The use of
CFT methods permits an algebraic formulation to be given for all of them. In some cases, we can prove
that there is a ground state, a Hamiltonian, and a CFT such that, for any number of particles,
all of these spaces are the same. For such cases, this resolves
several questions and conjectures: it gives a finite-size bulk-edge correspondence, and we can
use the analysis of functions to construct a projection-operator Hamiltonian that produces those
functions as zero-energy states. For a
model related to the ${\cal N}=1$ superconformal algebra, the corresponding Hamiltonian imposes
vanishing properties involving only three particles; for this we determine all the wavefunctions
explicitly. We do the same for a sequence of models
involving the $M(3,p)$ Virasoro minimal models that has been considered previously, using results
from the literature. We exhibit the Hamiltonians for the first few cases of these. The techniques
we introduce can be applied in numerous examples other than those considered here.
\end{abstract}

\pacs{PACS numbers: } 

\maketitle


\section{Introduction}
\label{intro}

\subsection{Overview}

In recent years there has been a considerable growth of interest in topological phases of matter (see
e.g.\ Ref.\ \cite{read12,TQCreview} for reviews). A topological phase of matter can be defined as an
equilibrium zero-temperature phase of a quantum system that has a gap in its energy spectrum above the
ground state, as far as local excitations far from any boundary are concerned. Topological phases may have
non-trivial properties (also called topological) that are unchanged by any (local) perturbation of the
(local) Hamiltonian unless it causes the system to pass through a phase transition. Model examples of
systems (consisting of a space of states, and a Hamiltonian) lying in a particular topological phase
provide existence proofs for the phase. Due to the gap in the spectrum, any sufficiently small
perturbation of the Hamiltonian by a local term will not destroy the gap. Hence, the solutions (even
partial solutions) of such model examples provide paradigms for those phases. Among the earliest
examples of topological phases and of the use of partially soluble models is the fractional quantum
Hall effect (FQHE) \cite{laugh,hald83}.

Historically, a standard approach has been the use of trial (or variational) states for the ground and
some of the excited states, even in the absence of a soluble Hamiltonian. In the FQHE, examples of such
states are usually described by a wavefunction that may readily be written down for arbitrary size
(particle number) in terms of multiplying some factors, performing integrals or sums, or other operations.
It is then hoped that these trial or ``nice'' wavefunctions have the topological properties of a
topological phase. In some cases, a Hamiltonian (or rather, a family of Hamiltonians, one for each system
size) is known for which the ground state is also known exactly for each size of the system. As generic
states are hard to specify concisely, such a state must have some special structure that enables it to
be specified in a simple way; thus it too must be a nice function.

Various approaches to obtaining trial wavefunctions have been used in the past. In the FQHE,
after Laughlin's
wavefunctions \cite{laugh}, there were generalizations using the hierarchy and composite particle
approaches \cite{hald83,halp,jain}, though some of these schemes possess considerable leeway when it
comes to
specifying definite functions. In this paper we will be concerned with another approach that developed
from Laughlin's states. It stemmed from the observation that the algebraic behavior of the
functions, which vanish as a power of separation as any two particles approach the same point, implies
that there is a ``pseudopotential'' interaction Hamiltonian acting in the space of lowest-Landau level
(LLL) states that annihilates both the Laughlin ground state and also the states containing any number of
fractionally-charged quasihole excitations \cite{hald83}. Thus the ``nice'' wavefunctions are actually
zero-energy eigenstates of a ``special'' Hamiltonian, and also exhaust all such states. The special
Hamiltonian consists of projection operators acting within the LLL; the coefficients of the projectors
are positive, so the Hamiltonian is  positive (semidefinite) as an operator within the LLL. Moreover,
the special Hamiltonian is usually translationally and rotationally invariant. It is also
local (short range), because it involves the low-degree behavior of the wavefunction in the differences
of particle coordinates as they tend to zero.

Another approach to obtain trial wavefunctions was developed by Moore and Read (MR) \cite{mr}, who
observed that the Laughlin wavefunctions, and some other previously-known trial wavefunctions, are
chiral correlation functions (so-called conformal blocks) in some two-dimensional conformal field theory
(CFT) \cite{bpz,dfms}. They extended this observation to propose some further trial states, and
the method used amounts to
a technique for the construction of an unlimited number of trial ground and associated quasihole
wavefunctions. Subsequently, it was observed \cite{gww,rr96} that the main example in MR, the MR or
Pfaffian state, is itself the zero-energy ground state of a special (i.e.\ projection operator)
Hamiltonian. This Hamiltonian is similar to those for the Laughlin state, in that it involves the way
in which the function vanishes as any {\em three} particle coordinates come to the same point, and so
consists of a three-body interaction. Using this Hamiltonian, explicit wavefunctions for any number of
quasiholes were constructed \cite{nw,rr96}. Other examples with similar structure have been found,
including notably
a sequence of further distinct topological phases in which the special Hamiltonians contain $k+1$-body
interactions, where $k=3$, $4$, $5$, \ldots \cite{rr99}; the preceding examples fit into the sequence as
the cases $k=1$, $2$.

Of the sets of ground and quasihole trial LLL wavefunctions that have been proposed, many of those that
can be expressed as functions, without use of integration over dummy variables (such as use of projection
operators, possibly from a larger Hilbert space such as those including higher Landau levels), can be
viewed as conformal blocks in a (known) CFT. Further, to our knowledge, {\em all} such sets of functions,
that appear acceptable as a space of ground and quasihole states in a topological phase and are precisely
the zero-energy states of a LLL special Hamiltonian, are conformal blocks. It should be emphasized that
even when all the zero-energy states of a special Hamiltonian can be found, the non-zero energy
eigenvalues generally cannot, and one does not know directly whether there is a gap in the spectrum of
bulk excitations over the ground state in the infinite size limit. Indeed, it has been argued that when
the wavefunctions are conformal blocks of either a non-rational or non-unitary CFT, any such Hamiltonian
must be gapless in the limit \cite{read09,read_chiral}. Consequently, in these cases the wavefunctions
do not represent the behavior of a topological phase, but may be at a critical point.

In this paper, the main goal is to provide tools to study systematically the short-distance behavior
of wavefunctions
(and related entanglement properties), use this to find special Hamiltonians, and analyze the spaces of
all zero-energy states of those Hamiltonians. The properties of the wavefunctions in these spaces
lends itself to an algebraic interpretation in general, and leads almost inevitably to the use of
MR constructions from a CFT to construct and analyze examples. In giving an overview of these ideas
and results to be presented in this paper, we will now divide them under the following headings.

\subsubsection{Zero-energy subspaces for a given Hamiltonian}

LLL wavefunctions for bosons are essentially
symmetric polynomials in the complex coordinates $z_j=x_j+iy_j$ for the $j$th particle with Cartesian
coordinates $x_j$, $y_j$ in the plane. From here on, when we refer to functions, it is such symmetric
polynomials that we mean. The phrase ``short-distance behavior'' means a Taylor expansion in some
selected particle
coordinates as they all tend to some value; hence we usually organize the spaces by degree of the
polynomials, beginning with the lowest degree. (We concentrate on bosons; wavefunctions for fermions may be
obtained by multiplying by $\prod_{i<j}(z_i-z_j)$, and there are related Hamiltonians.) Given a
translationally- and rotationally-invariant projection operator Hamiltonian that consists of interactions
among at most $N_{\rm max}$ particles, one can in principle determine all of the zero-energy states for
each number of particles $N=1$, $2$, \ldots in the plane. We call the spaces of these states the
{\em zero-energy (sub-)spaces} of states or wavefunctions; they are subspaces of the space of all LLL
states for given $N$, and usually only of interest when they are {\em proper} subspaces. We also
describe the wavefunctions as having
``allowed'' behavior (one might also describe them as ``satisfying'' the constraints that they are
annihilated by the projectors in the special Hamiltonian). In the literature, these are often called
spaces of edge excitations, though they can only be considered as true edge excitations in the case
when there is a gap for bulk excitations \cite{wen_edge,milr}. (One may also consider other
geometries, including the sphere
with a given number of flux quanta emerging from it, which lead to related results. For a gapped bulk
phase on the sphere, the zero-energy states may be viewed as quasihole states.) In the body of this
paper, we carry out the analysis of the zero-energy subspaces for the plane and the sphere explicitly
in some examples.

\subsubsection{Entanglement subspaces for a given
ground state}

In analyzing the zero-energy states, the spaces for $N\leq N_{\rm max}$ play a special role. If these
are known, then the zero-energy wavefunctions for $N>N_{\rm max}$ particles must be found by some process
of ``patching together'' the zero-energy wavefunctions for sets of $N_A\leq N_{\rm max}$ particles.
Put another way, if the former
are decomposed into wavefunctions for a subset of $N_A\leq N_{\rm max}$ particles and for the remainder,
the wavefunctions for the $N_A$ particles must lie in the zero-energy subspace. Indeed, the same
is true for the decomposition for {\em any} value of $N_A\leq N$. For the ``ground state'' in
the zero-energy subspace for any $N$, that is the (usually unique) wavefunction of lowest total degree
in the $z_i$'s (lowest angular momentum), we term the space of functions that actually occur when it
is decomposed into functions of $N_A$, times functions of the remaining $N-N_A$, coordinates an
{\em entanglement subspace}, because of its relation (discussed below) to entanglement properties
of a ground state wavefunction. From the preceding remarks, it is clear that the entanglement subspaces
for any $N_A$ (for the ground state at any $N\geq N_A$) must lie within the zero-energy subspace for
$N=N_A$ particles. [One could of course do the same
for any zero-energy state; we sometimes refer to these also as entanglement subspaces.] In the body of
this paper, we analyze the relation of the zero-energy and entanglement subspaces, in some examples.

\subsubsection{Inverse problem: finding a special Hamiltonian for given zero-energy subspaces}
\label{sec:compat}

If there are some polynomials ``missing'' from the entanglement subspace for some value of $N_A$, and
some degree $d$, for all values of the total number $N$, then a projection operator onto that function
can be added to the Hamiltonian, and it will still annihilate the wavefunctions. More precisely,
the projection must be onto some part of the orthogonal complement of the ``allowed'' functions.
(The orthogonal complement and the construction of projection operators involve the use of the
quantum-mechanical inner product on the wavefunctions of our system.) The special Hamiltonian with
which we began
must itself consist of such projectors, and one would like to use one with a minimal number of terms.
Then we also have the inverse problem: if we are given some spaces of wavefunctions, for all
particle numbers $N$, and we wish to obtain a special Hamiltonian for which these are precisely the
zero-energy subspaces, then such an analysis of the entanglement subspaces
will ultimately lead to the solution, if one exists. A necessary condition for existence is that the spaces
of functions should obey the compatibility condition mentioned in the preceding paragraph. That is,
all the entanglement subspaces for $N_A$ particles in {\em any} $N$-particle function in the given
space ($N_A<N$) should be a subspace of the given space of functions for $N_A$ particles.

\subsubsection{Spaces of amplitudes from a CFT}

In particular, as spaces of functions, we can begin with conformal blocks in a MR construction.
This produces spaces of functions, which we may also term ``spaces of amplitudes'', that play the
role of the zero-energy subspaces; the spaces depend on the CFT used. In these constructions, the
necessary compatibility conditions are automatically satisfied, because the short-distance analysis
of entanglement subspaces is related to the operator-product expansion in CFT, which has such behavior
\cite{rr99}. We may attempt to find a corresponding special Hamiltonian for which such spaces of
amplitudes are the zero-energy subspaces.

\subsubsection{Entanglement}

The analysis so far has led to issues of entanglement.
For entanglement in general in a many-particle state, one effectively divides the system into one part with
$N_A$ particles, and the remainder with $N_B=N-N_A$ particles. One may trace over the states of the $N_B$
particles to obtain a reduced density matrix, and then analyze its eigenvectors and eigenvalues
\cite{kp,lw}. Thus
essentially, the space in which the eigenvectors lie (the entanglement subspace) is that spanned by the
wavefunctions of the $N_A$ particles that occur in the ground state, just as in the preceding discussion
(we will discuss different ways of defining the bipartition of the system in more depth later). If the
partition respects rotation invariance about the origin, then the states can be classified according to
their angular momentum, which is the same as their degree as polynomials. In some ways of defining the
bipartition to obtain entanglement, a range of values of $N_A$ must be considered also. Several ways of
bipartitioning for LLL states have been considered in the literature; for our purposes, which relate to
special Hamiltonians, the most appropriate is particle partition \cite{schoutens1}, which
preserves translational, as well
as rotational, symmetry. (Many studies of entanglement use the sphere; in that geometry particle
partition has the full rotation symmetry.) When studying
entanglement, one is often interested in universal behavior as $N_A$ (or its ``typical'' value, if
there is a range) tends to infinity (either subsequent to, or at the same time as, $N\to\infty$).

An important idea is that in LLL trial states for a topological phase in the FQHE (in fact, for
those obtained  as conformal blocks/amplitudes from a MR construction), the multiplicities of
the states in the decomposition at sufficiently large angular momentum and $N_A$---that is, the
dimensions of the entanglement subspaces---approach those of
the edge states \cite{lh}, which can sometimes be identified as the zero energy states of a special
Hamiltonian. For states given by the MR construction as conformal blocks, these multiplicities
(in the limit) are those of the vacuum sector of the CFT. More generally, in that construction,
the entanglement subspaces must lie within the spaces of amplitudes, because of the compatibility
conditions already mentioned.
We will show, by general algebraic arguments, that the dimension of the space of {\em amplitudes}
obtained by the MR construction {\em does} agree with the vacuum sector of the CFT used as $N\to\infty$
(in an appropriate way). We also show that, in the limit $N_B\to\infty$ with $N_A$ fixed, the
entanglement subspaces contain {\em all} the same functions as those in the spaces of amplitudes for
$N=N_A$, at {\em any finite} value of $N_A$ as well, provided the correct CFT is used to
construct amplitudes. Hence in particular, their dimensions are equal. This is a finite size ($N_A$
finite, though $N_B$ went to infinity) version of a bulk-edge correspondence,
if the spaces of amplitudes are viewed as edge states. The ``correct'' CFT is one
without any ``singular'' vectors, or alternatively can be obtained from the given CFT by a process
of setting the singular vectors to zero, to form the quotient space \cite{bpz,dfms}.
(Singular vectors can only occur in non-unitary CFTs.) The equality of the {\em entanglement} subspaces
with the spaces of amplitudes does
{\em not} hold for general $N_A$ as $N_B\to\infty$ if there are singular vectors in the CFT that was
used to find the amplitudes, even though the ground state wavefunctions, that result from the
construction using the CFT instead of its quotient, are the same. We give examples of this
phenomenon.

\subsubsection{Algebraic techniques}

We develop an algebraic language to aid in all of this analysis. The space of all symmetric
polynomials in $N_A$ variables is a vector space, and can be viewed as the Hilbert space of LLL states
of bosons. But it can also be viewed as an algebra, because not only are sums and complex multiples
of symmetric polynomials again symmetric polynomials in the same variables, but so are products
of them. Further, this algebra acts in an obvious way, by multiplication, on the vector space
of symmetric polynomials, which can then also be viewed as a ``module'' over
(or in less fancy language, a representation of) this algebra.

It is very convenient if the zero-energy subspaces, which are subspaces of symmetric polynomials,
are themselves modules over the symmetric polynomials---that is, if the subspace is closed under
multiplication by symmetric polynomials. That is because, as we will see when explicitly finding
allowed functions,
this property allows us to combine factors for different subsets of the particles that satisfy the
conditions on allowed functions separately, secure in the knowledge
that they will also do so when combined as $N$ particle wavefunctions. This property holds in
previously known examples, and in those we analyze here.  If the special Hamiltonian
consisted of an arbitrary set of projectors (i.e.\ if the allowed functions did not form a module over
the symmetric polynomials), it would be much harder to construct the spaces of zero-energy many-particle
states, and it would not be clear if any such states even exist. We further require that our
Hamiltonians be short-range interactions, and that they be translationally and rotationally
invariant. Rotational invariance implies that angular momentum, or the degree of the polynomials,
be a good quantum number; we may classify functions by degree. Translational invariance implies
that the allowed behavior is mapped into itself by the application of the center of mass variable
$\propto \sum_i z_i$ and $\sum_i \partial/\partial z_i$ (acting on the polynomials).

Consequently, when searching for special Hamiltonians, we now have an algebraic problem, involving the
study of translation-invariant modules
for the algebra of symmetric polynomials. While this problem can be solved
completely for the case of $N_A=2$ particles (see Section \ref{sec:sub_ham} below), it is much more
complex for more than two particles. To make
progress, motivated by the MR construction using CFT, we can also enlarge the algebra of symmetric
polynomials by introducing further operators which form part of the action of the Virasoro algebra on the
wavefunctions. This algebraic problem can also be studied, without further use of CFT. Because the
requirement of being a module over the larger algebra is more stringent, there are fewer solutions to be
found, and they potentially arise from conformal blocks. Conversely, given such a Hamiltonian, we may
find its zero-energy states, as we will do in several examples.

Further motivation for confining the search to spaces with such an (extended) module structure
can be obtained
from arguments in Ref.\ \cite{read_chiral}. It was argued there that for a special Hamiltonian that has
a gap in the bulk spectrum (and no other assumptions), the space of zero energy (or edge) states in
the limit of large sizes forms a chiral conformal field theory, with an action of both the Virasoro and
U$(1)$ current algebras. However, it is not clear if all the operators that generate these
algebras (or of the subalgebras that we use in this paper) must have the one-body form that we
encounter below, even though certain of them, which relate to translations and rotations of all
the particles, do. (The arguments of Ref.\ \cite{read_chiral} also lead us to expect that the action
of the current and Virasoro generators must be local in position space; it may be that this further
condition implies that the one-body form must hold.) Nonetheless, such a form would seem natural,
and the assumption that there is this module structure, which we will make even for cases that we
don't expect to be gapped in the bulk, and in finite size, will ensure that when the bulk is gapped
(the cases of most interest to us anyway), this physics will emerge.

In the spaces of amplitudes produced by a MR construction, the entanglement subspaces automatically
form a module over the symmetric polynomials---corresponding to the presence of a U$(1)$ current algebra in
the CFT---and over the part of the Virasoro algebra mentioned above, and perhaps also over some larger
algebra, depending on the CFT considered. In this set-up, the ``missing'' polynomials (corresponding to
the orthogonal complement of an entanglement subspace) are related to a finite-size version of the
presence of singular (also called null) vectors in the underlying Verma modules of the CFT, a
topic familiar in CFT. If we understood these orthogonal complement spaces, we would be able to
find projection operators that could be included in a special Hamiltonian whose zero-energy spaces
contain the given spaces of amplitudes. We would then be left with the problem of finding a
{\em minimal} Hamiltonian such that the latter two sets of spaces actually coincide.

\subsubsection{Superconformal and $M(3,p)$ examples}

To illustrate the preceding ideas in a concrete manner, we introduce a family of examples.
The original motivation for this work was to find examples of special Hamiltonians for which the
zero-energy states are precisely the spaces of amplitudes obtained from a unitary rational CFT; a
guiding example was the tricritical Ising state \cite{read09} for which the ground state wavefunction
is known \cite{simon09}. Unfortunately we have not succeeded in finding a special Hamiltonian for
this example. But Ref.\ \cite{simon09} gives a family of ground state wavefunctions, one for each
of a continuous family of special Hamiltonians, and those are labeled by one parameter $c$ and involve
three-body interactions only. These wavefunctions are obtained from the MR construction
using the ${\cal N}=1$ superconformal field theories \cite{estienne1}, and include the tricritical 
Ising ground
state at one special value of $c$, the central charge. (In general, these SCFTs are neither unitary
nor rational.) The filling factor in these states is $\nu=1/3$.

In Section \ref{scft} of this paper we analyze the zero-energy spaces for this family of
special Hamiltonians (with no use of CFT), and find their dimensions. We also show that this
family of models based on ${\cal N}=1$ SCFT, which are expected to have gapless bulk excited
states in the limit, are adjacent to a MR phase of bosons at $\nu=1/3$, consistent with their
being on a phase boundary. In Section \ref{sec:amps}, we show that (except for $c=0$, which has larger
degeneracy) in the case of the plane geometry these dimensions agree in the
infinite size limit with those of the generic ${\cal N}=1$ superconformal field theory (SCFT)
construction. The amplitudes constructed from the SCFT are guaranteed to be zero-energy states, and in fact
we can eventually prove that (for $c\neq0$) the zero-energy subspaces for any finite $N$ are identical
with the spaces of amplitudes of that SCFT for the same $N$ (that is, not only the ground
state wavefunction, but all wavefunctions in these spaces). Hence these spaces are
also equal (at generic values of $c$, as we discuss in a moment) to the entanglement subspaces at
$N_A$ equal to this value $N$ (as $N_B\to\infty$); the functions in, and the dimensions of, these spaces
are then all known. These arguments are very general and can be applied to many other MR constructions,
whenever the zero-energy states can be constructed and the dimensions of the spaces agree with those of
the vacuum of the CFT in the large-size limit.

At the special point corresponding to the tricritical Ising model, these
generic dimensions are strictly larger than those for the spaces of amplitudes in the latter.
This is because at this value of $c$, the modules of the generic ${\cal N}=1$ SCFT (which will
be precisely defined later) contain singular vectors. The quotient space, obtained by setting the
singular vectors to zero (or ``modding them  out''), yields a module that has no remaining
singular vectors. The same phenomenon occurs at a
countable infinity of other ``rational'' values of $c$, and is connected with the ${\cal N}=1$ SCFT
minimal models. In general, the SCFT minimal models (like the Virasoro minimal models) are rational
but not unitary, however, there is a subset that are also unitary; among the latter, the tricritical
Ising theory is the non-trivial theory of lowest $c$.

These facts have several consequences. First, at these rational values of $c$, the entanglement
subspaces of the ground state, which as we will show always agree with the spaces of amplitudes
in the MR construction using the {\em quotient} or minimal model SCFT, have lower dimensions than those
at generic $c$. This shows, as mentioned above, that the correspondence of dimensions of the
entanglement and zero-energy spaces does not always hold. Hence, as $c$ varies, some
pseudoenergies in the entanglement spectrum must move off to infinity at these values of $c$, and this
will occur even in finite size (sufficiently large, though how large will depend on the $c$ value
involved).

Second, at the rational values of $c$, the ``missing'' polynomials in the entanglement subspaces for
some values of $N_A$ allow us to add corresponding terms to the special Hamiltonian which then still
annihilates the ground state. These terms depend on $c$ and do not annihilate the ground state at
generic $c$.

In order to determine, for a particular rational value of $c$, whether addition of such terms in
the Hamiltonian produces zero-energy states that coincide with the spaces of amplitudes in the
MR construction using the quotient SCFT at the same $c$, one must analyze the zero-energy states, and
then if they are not the same as the spaces of amplitudes, add further terms to the Hamiltonian.
This seems difficult to carry through
in general. However, a remarkable paper by Feigin, Jimbo, and Miwa (FJM) \cite{fjm} (whose
algebraic approach is very similar to ours, though without the physical applications in view here),
shows in effect that for one rational value of $c$, there is a three-body special Hamiltonian whose
zero-energy subspaces are precisely the spaces of amplitudes, for any particle number $N$. In fact,
their analysis applies to a sequence of MR constructions using Virasoro minimal models $M(3,p)$ [$p$
is a positive integer, and not divisible by $3$; $p=4$ gives the MR state, and $M(3,5)$ is the
Gaffnian \cite{srcb}]. These spaces of amplitudes have been discussed previously in relation to the
FQHE, but
the existence of a special Hamiltonian in these models was apparently overlooked (except for $p=4$,
$5$, for which the Hamiltonians were known). We give the Hamiltonians for several of these in explicit
form, and find all their zero-energy wavefunctions for all $p$ for both the plane and sphere geometries;
the multiplicities of these reproduce results in FJM.

\subsubsection{Structure of the paper}

The structure of this paper is as follows.
In the remainder of this section, we review some of the earlier work that is relevant for our paper,
and describe the issues in greater detail. The basic issues are dual: special Hamiltonians and
subspaces of ``nice'' trial functions. This duality leads to a ``chicken and egg'' problem,
in that a special Hamiltonian (the chicken) produces spaces of nice zero-energy states (the eggs),
while in turn spaces of states that satisfy the compatibility conditions can be used to produce
Hamiltonians that annihilate them. It is not obvious with which of these two types of objects one
should begin. But as both types of analysis hinge on the entanglement subspaces, which correspond to
(possible) terms in a special Hamiltonian, we begin with those.

In Section \ref{scft}, we consider the wavefunctions determined by the (continuous family of)
``vanishing conditions'' on the relative variables for any three particles, which is related to the
${\cal N}=1$ SCFT. The methods make no use of CFT, and give the complete, linearly-independent states
for both the plane and sphere geometries. The count of states on the plane is interpreted later. We
also argue that this model sits at a phase boundary adjacent to a MR phase.

In Section \ref{sec:amps}, we begin to use methods from CFT. We describe features of the MR construction
in the operator language of vertex operator algebras (VOAs). We prove a Theorem that implies that the
functions obtained from the MR construction span a space isomorphic to the VOA. We show that the
entanglement subspaces of the ground state in particle partition span all the amplitudes in the limit
as $N_B\to\infty$ if the CFT is non-degenerate, but in degenerate cases span a smaller space. We show
that in CFT-related examples in which the count of zero-energy states of a special Hamiltonian can
be carried out, these numbers and the wavefunctions themselves agree with the amplitudes even in
finite size $N$.

Finally, in Section \ref{sec:M3p}, we show in an example how when the entanglement subspaces are smaller
than the space of amplitudes of a CFT (or zero-energy states of a special Hamiltonian), because the CFT
contains singular vectors, terms can sometimes be added to the special Hamiltonian such that agreement
is restored with the quotient CFT. The examples in this section use the $M(3,p)$ minimal models.

Appendix \ref{app:c0} deals with a special case postponed from Section \ref{scft}.
The proofs of two theorems stated in Section \ref{sec:amps} are given in the other Appendices;
that of Theorem 1 in
Appendix \ref{app:lim}, and that of Theorem 2 in Appendix \ref{app:thm2}.

\subsection{Decomposition of trial wavefunctions}
\label{sec:decomp}

We consider the example of the Laughlin-Jastrow wavefunction \cite{laugh}
\be
\Psi_{\rm L}(z_1,\ldots, z_N)=\prod_{i<j}(z_i-z_j)^Q,
\ee
where we have dropped the ubiquitous factor that is a function of $|z_i|^2$ (which is Gaussian for the
plane \cite{laugh}, rational for the sphere \cite{rr96}) as they do not enter the argument here. The
degree in each $z_i$ is $N_\phi$ (the number of flux quanta covered by the particles),
\be
N_\phi=QN-Q.
\ee
If we include quasihole factors $\prod_i(z_i-w_k)$ for each of a set of quasiholes at positions $w_k$
($k=1$, \ldots, $n$), then we know that each quasihole factor can be expanded in elementary symmetric
polynomials
\be
\prod_i(z_i-w_k)=\sum_{m=0}^N e_m(-w_k)^{N-m},
\ee
where
\be
e_m=\sum_{i_1<i_2<\cdots i_m}z_{i_1}z_{i_2}\cdots z_{i_m}
\ee
($m=1$, \ldots, $N$) are the elementary symmetric polynomials in $N$ variables, and for convenience we
also define $e_0$ by $e_0=1$.
The elementary symmetric polynomials generate the algebra of all symmetric polynomials in $N$, and do
so {\em freely} (there are no linear relations between products of elementary symmetric polynomials).
Then expanding the product of quasihole factors in powers of the $w_k$'s
produces symmetric polynomials, and by using sufficiently many quasiholes we can eventually produce all
the symmetric polynomials. [As the functions are symmetric under permutations of the $w_k$s, this
expansion can be viewed as a double expansion in symmetric polynomials in the two sets of variables].
These functions which contain the Laughlin-Jastrow factor times arbitrary symmetric polynomials can
be viewed as representing the edge excitations of the Laughlin disk.

The Laughlin wavefunction itself has a somewhat similar form as the quasihole factors. Thus if we divide
the particles into two groups, say $\{1,\ldots,N_A\}$ and $\{N_A+1,\ldots,N\}$ (and let $N_B=N-N_A$),
then we may write the function in the form
\bea
\Psi_{\rm L}(z_1,\ldots,z_N)&=&\sum_r \psi_{Ar}(z_1,\ldots,z_{N_A})\non\\
&&\,{}\times\psi_{Br}(z_{N_A+1},\ldots,z_N),\label{prod_decomp}
\eea
where $r$ is an arbitrary index and the functions $\psi_{Ar}$ and $\psi_{Br}$ are symmetric in their
arguments. If $\Psi_{L}$ were replaced by a generic state (symmetric polynomial), the functions
$\psi_{Ar}$ obtained in this way would span the set of
symmetric polynomials times $\psi_{\rm L}$ in $N_A$ variables, at least in low degree (clearly the degree
of $\psi_{Ar}$ must be less than $NN_\phi/2$, the degree of the state considered). It is natural to wonder
if this is the case for the Laughlin state. There are several variants of
this question (in each case, initially with $N_A$ fixed): (i) in one version, we consider all polynomials
obtained as $N_B$ increases to infinity; (ii) in a second, $N_B$ is instead fixed, and it follows that
the highest degree in any $z_i$ with $i\in {\cal A}$ is bounded by $N_\phi$ [version (i) corresponds
to the limit of this as $N_B\to\infty$]. In one further version, (iii) we require that $N_B$ be fixed,
and also that in the functions $\psi_{Ar}$ the degree in each $z_i$ in $\{1,2,\ldots,N_A\}$ be less than
or equal to some bound less than $N_\phi$, while the degree of $\psi_{Bm}$ in each $z_i$ in
$\{N_A+1,\ldots,N\}$ must be strictly greater than the same bound, similarly. In this
case, an explicit symmetrization over permutations of the coordinates is required in order to yield a
valid decomposition of $\Psi_{\rm L}$ in eq.\ (\ref{prod_decomp}) (it is not required in the others
preceding). In each version, one may finally consider different $N_A$, and take the limits as
$N_A\to\infty$. These decompositions can be applied in the same way to other trial wavefunctions also,
and the limits can be studied as there is such
a wavefunction for each $N$ (or for each $N$ divisible by some $k>0$).

These decompositions of the Laughlin wavefunction are equivalent to definitions of entanglement that
have been considered recently. For entanglement, we assume that the Hilbert space $\cal H$ has a
tensor product form,
\be
{\cal H}={\cal H}_A\otimes{\cal H}_B,
\ee
at least within each sector of fixed $N_A$. If we represent vectors
in each space by wavefunctions, then we can decompose a function $\psi$ accordingly as
\be
\psi=\sum_r \psi_{Ar}\psi_{Br},
\ee
where $\psi_{Am}$ ($\psi_{Bm}$) belongs to ${\cal H}_A$ (${\cal H}_B$).
The question is to characterize the degree and nature of the {\em entanglement}, that is the extent to
which $\psi$ is not a tensor product. (This is rather too naive for systems of many identical
particles, which because of symmetrization do not have a product form. However, once some particles have
been assigned as part $A$ and others as $B$, the overall symmetry of the state is broken, and the product
form applies.) One version for QH states
corresponds to further restricting the single-particle LLL Hilbert space according to the angular momentum
of each particle, which corresponds to the degree in each $z_i$, so that some values appear only in part
$A$, and the remainder in part $B$. This ``orbital'' entanglement \cite{schoutens1,lh} corresponds to
the third version above. The entanglement subspaces spanned by
the $\psi_{Ar}$ for some trial wavefunctions were found to be smaller than the spaces of all symmetric
polynomials of the correct degree---this is essentially the phenomenon of the ``entanglement gap''
discovered in Ref.\ \cite{lh}. Other studies have considered ``particle'' entanglement
\cite{schoutens1}, corresponding to the second version above. (A more formal extended definition
of particle partition, in particular, was given in Ref.\ \cite{drr1}.)
It has been conjectured that the dimensions of the entanglement subspaces in particle partition
on the sphere agree with the counting of quasihole states for $N=N_A$, $N_\phi$, provided $N_B$ is
sufficiently large \cite{srb}, however this does not seem to have been proved to our knowledge.
Another version
of the entanglement spectrum is obtained by considering a bipartition in position space
---the real-space entanglement spectrum \cite{drr1,sterd,rss,drr2}, which we will not describe
further at the moment. The limit that has usually been considered for orbital and for real-space
entanglement is that in which $N\to\infty$ with the ratio $N_A/N$ fixed. The first version
above ($N_B\to\infty$ with $N_A$ fixed) has not been considered as much.

As mentioned above, in Section \ref{sec:amps} of this paper, we obtain detailed statements about
the entanglement spaces in particle (and also real-space) entanglement, in the first version above,
the limit $N\to\infty$ with $N_A$ fixed. We believe this is the simplest, or most basic, limit of all.
That is, we can show that after taking $N_B\to\infty$, the functions $\psi_{Ar}$ span a space isomorphic
to the space of amplitudes in $N_A$ variables, where the CFT used to construct the amplitudes is a quotient
of the CFT used in the construction, obtained by setting any singular vectors to zero. Moreover, as
$N_A\to\infty$, the space of amplitudes in a MR construction
becomes isomorphic to the space of states in the vacuum representation of the same CFT.  We note
that there has been recent progress in establishing such relations for real-space and particle
partitions of trial wavefunctions \cite{drr2}, with which the present result has some overlap. However,
the arguments presented there made use of generalized screening behavior to obtain results about the
inner products of the wavefunctions, which yields much more information about the entanglement
spectrum (not only dimensions of subspaces) than we obtain here. Such behavior is not expected to hold
for trial wavefunctions obtained from non-unitary or non-rational CFTs. By contrast, the method used
in the present paper is purely algebraic in that the quantum mechanical inner product on the
wavefunctions is not used, and while it gives less information, it works for all cases, including
non-unitary and non-rational ones.

\subsection{Subspaces and Hamiltonians}
\label{sec:sub_ham}

As we know, the entanglement subspaces are the key to searching for a special Hamiltonian for a given
set of wavefunctions.  Several works have addressed this problem, beginning from Ref.\ \cite{hald83}.
Recently, a start has been made on systematically studying the possible special Hamiltonians for
more than two-body interactions, and finding their zero-energy many-particle states \cite{src}.

It will be useful to describe the spaces of symmetric polynomials in $N$ variables in more detail.
Because the algebra of symmetric polynomials is freely generated by the elementary symmetric polynomials
(or, in fact by some other alternative sets of $N$ generators), the number of symmetric polynomials
in each degree can be found easily. Generally, for a vector space $V$ of symmetric polynomials that
have multiplicities (dimensions) $b_d$ in each degree $d$, we can define the character,
\be
{\rm ch}_q\,A=\sum_{d=0}^\infty b_d q^d
\ee
where $q$ is an indeterminate (the series is a formal power series, which means that its convergence is
not of interest). For the space $\Lambda_{N}$ consisting of all symmetric polynomials in $N$
variables, the character is
\be
{\rm ch}_q\,\Lambda_{N}=\prod_{n=1}^{N}\frac{1}{(1-q^n)}=\frac{1}{(q)_N},
\ee
where we have introduced the compact and standard notation,
\be
(q)_m=\prod_{n=1}^m(1-q^n).
\label{q_m}
\ee
Because the Hamiltonian should be translationally (and rotationally) invariant, it is useful to
decompose the possible behavior for $N$ particles into the center of mass and some ``relative'' or
``internal'' wavefunction. The center of mass coordinate is $Z=\sum_i z_i/N=e_1/N$, and the center
of mass wavefunctions are spanned by $Z^m$, $m=0$, $1$, $2$, \ldots. The center of mass angular
momentum can be raised by multiplication by $Z$, and lowered by applying $\partial/\partial
Z=\sum_i \partial/\partial
z_i$. (In these expressions, the sums over particles are from $i=1$ to $N$. We can treat the derivatives
as acting on the polynomials part of the wavefunction only \cite{gj}). For the internal part,
coordinates can
be defined, and a space of functions of these coordinates orthogonal to the center of mass variable can be
constructed. However, there is no particularly natural definition for these. In any case, the functions
in the internal variables must be annihilated by $\partial/\partial Z$, which means that the polynomials
are translation invariant---they are functions of differences $z_i-z_j$ only. The space of general
wavefunctions is thus spanned by $Z^m$ times polynomials of the differences of coordinates.
(The space of translationally-invariant symmetric polynomials can be freely generated by a modified
set of $N-1$ elementary symmetric polynomials, given in Ref.\ \cite{src}.) For two
particles, the latter have the simple form
\be
(z_1-z_2)^{m'}
\ee
where again $m'=0$, $1$, \ldots, and now $m'$ must be even for bosons (symmetric polynomials) and odd
for fermions. There is only one polynomial for each even (resp., odd) degree. This is not typical; for
$N>2$, the number of linearly-independent polynomials in each degree increases gradually.
The space of polynomials in $Z$ has character $1/(1-q)$, and so the character of the space of
wavefunctions for the internal motion is
\be
\prod_{n=2}^{N}\frac{1}{(1-q^n)}.
\label{charint}
\ee
For example, for $N=3$ particles, we have
\be
\frac{1}{(1-q^2)}\frac{1}{(1-q^3)}=1+q^2+q^3+q^4+q^5+2q^6
+q^7+2q^8+\ldots,
\label{ch_N3}
\ee
and the coefficients have the pseudoperiodic property that
if the degree increases by $6$, the coefficient increases by $1$.

Following earlier examples, we may attempt to describe some spaces of polynomials by specifying the
behavior that is ``allowed'' for members of the space as some numbers $N_A$ of the particles come
together; these are sometimes termed ``vanishing conditions''. As discussed above, we choose to require
the spaces to be modules over the algebra of symmetric
polynomials. It is not clear if this property is really necessary. Physically, multiplying by a symmetric
polynomial generates an excitation of the charge sector only.

For $N_A=2$, the translation-invariant modules over the symmetric polynomials can be completely
determined. There is only one generator, $(z_1-z_2)^2$, so any such module is spanned by the even powers
$(z_1-z_2)^{m'}$ as above, for $m'=m$, $m+2$, \ldots, times a power $Z^M$, $M=0$, $1$, \ldots . Thus the
distinct translation-invariant modules are labeled by $m=0$, $2$, \ldots, and the elements of each such
module can be generated by $(z_1-z_2)^m$ by multiplication by symmetric polynomials. These correspond to
the distinct special or pseudopotential Hamiltonians for bosons that produce the Laughlin ground,
quasihole and edge states as zero-energy states \cite{hald83} (it is similar for fermions, if we
consider modules consisting of antisymmetric polynomials). Cases that include conditions on more
than two particles are much richer. They include the MR state for bosons at $\nu=1$, which arises from
requiring the polynomials to vanish if any three coordinates coincide. This corresponds to the degree $0$
term in the above series, which represents the dimension of the relative-coordinate space of symmetric
polynomials. Because there is no relative term at degree $1$, the functions vanish quadratically as the
three coordinates come together (in any fashion, e.g.\ first two, and then the third, or with any
fixed ratio of pairwise separations). The more-recently proposed Gaffnian ground state, and its edge
and quasihole states, arises if we require the functions to vanish as degree $3$ when three coordinates
approach the same point \cite{srcb}. Both of these spaces of allowed behavior form a module over the
symmetric polynomials.

In the examples considered in Ref.\ \cite{src}, the allowed behavior was defined by requiring that
all relative motion with degree less than some bound was forbidden, and all greater was allowed.
We note that this
automatically produces a module over the symmetric polynomials. However, not all those examples led to
truly ``nice'' behavior, with a unique ground state (lowest total degree polynomials) for $N$ particles,
and moderate growth of the number of allowed or zero-energy states at higher degree. Based empirically on
the evidence of past cases, such behavior would be highly desirable when searching for new model QH phases
of matter. Other than the recent Gaffnian \cite{srcb}, no such nice spaces of functions were identified
in that work that had not been found previously. In further work \cite{srr}, some four-body
interactions were studied, and related to a MR construction using $S_3$ CFTs 
\cite{estienne1}. The methods developed in the present paper can be applied to that case also.

At the higher degrees, for example degree $6$ or larger than $7$ for the three-particle case above, there
are more possibilities. We may choose to allow, for three particles, any degree greater than $6$, and only
a one-dimensional subspace of the space in degree $6$. The choice of a subspace means that there is now
a {\em family} of spaces of functions (a one-parameter family in this example), or a family of
Hamiltonians, each producing one of the spaces. We will study this example in detail in this paper.
We emphasize that the possibility of such a parameter or parameters
was not included in the original versions of the ``thin-torus-limit'', ``pattern-of-zeroes'', and
``root state'' approaches \cite{bkwhk,sl,rhsl,wenwang,bh1,bh2,bh3}. But the objects under study are
vector spaces, which allow use of such linear combinations
(when the dimension is greater than one), not discrete objects such as sequences of integers. The latter
cannot represent general {\em vectors} in a vector space, though they can represent {\em basis} vectors,
such as the monomial basis for many-particle states in the lowest LL. We should add that later versions of
some of these approaches have gradually included more terms in the Hamiltonian, corresponding to
subleading terms in the thin-torus limit, and the explicit use of CFT constructions rather than
direct ``classification'' of spaces of polynomials in the pattern of zeroes approach.

\subsection{Moore-Read construction}

In the MR construction \cite{mr}, trial ground state wavefunctions are constructed as correlation
functions in a two-dimensional CFT. More precisely, the functions are conformal blocks, or chiral parts of
correlation functions, which depend only on $z$, not on $\overline{z}$, and take the general form
\be
\langle 0|{\cal O}_{-N}\prod_{i=1}^N a(z_i)|0\rangle
\label{ground}
\ee
for ground state, where $|0\rangle$ is the vacuum of the CFT, and
\be
a(z)=e^{i\varphi(z)/\sqrt{\nu}}\psi(z),
\ee
where $\varphi(z)$ is a free chiral scalar field, and $\psi(z)$ is a Virasoro primary field with
Abelian statistics from some CFT \cite{bpz,dfms} (more precise requirements will be explained later).
The coefficient of $\varphi$ in each exponential is a U$(1)$ charge. ${\cal O}_{-N}$ is an operator of
charge $-N$ (so that overall neutrality is satisfied). In MR, ${\cal O}_N$ was $e^{-i\int_{\cal D} d^2z
\rho}$ for some uniform density $\rho$ and domain $\cal D$, such that it carries total charge $-N$.
By including additional fields inside the chiral correlator, it is possible to produce
wavefunctions describing quasiholes as well; we will not require the explicit form of these in this paper.

The operator product expansion (ope), essentially the short-distance expansion of correlation functions,
plays an important role in CFT. The properties of the operators $a$ as their coordinates become equal
determines the algebraic structure of the theory. This information is then related to the entanglement
subspaces already discussed. We will provide very detailed connections between the two later in this
paper.

There is a completely general duality, explained in Section \ref{sec:amps} below, that allows {\em any}
spaces of symmetric polynomials that are characterized by ``vanishing conditions'' (or equivalently as
the zero-energy spaces for some special Hamiltonian) to be constructed in a similar form (again see
Ref.\ \cite{fjm}, and also the earlier \cite{fs}. In an
informal account, we introduce an infinite set of operators $\xi_{-n}$, $n=0$, $1$, \ldots,
all of which commute, and let $\xi(z)=\sum_{n>0}\xi_{-n}z^n$; they form an algebra $\cal A$.
We suppose that there is a vacuum $|0\rangle$. Then the vector space spanned by the vectors
$\prod_{i=1}^N \xi_{-n_i}|0\rangle$ is dual
to the space of symmetric polynomials in $N$ variables (much as if each $\xi_{-n}$ is a creation operator
for a boson in the $n$th orbital in the LLL). But now we may also impose local, $z$-independent
algebraic relations among the $\xi(z)$'s, such as $\xi(z)^2=0$, or $\xi(z)\partial\xi/\partial z=0$.
In general, each relation can be a linear combination of products of $\xi$ and its derivatives all
evaluated at $z$, and must be homogeneous both in $\xi$ and in derivatives. Clearly, if such a relation
is multiplied by any number of $\xi(z_i)$ at arbitrary positions, or if linear combinations are formed,
other valid relations are obtained---thus the combinations that are to be set to zero form an ideal
$\cal I$ in the algebra $\cal A$ with which we started. Then the symmetric polynomials that are dual
to the quotient ${\cal A}/{\cal I}$ acting on $|0\rangle$ must obey vanishing conditions as some number
of $z$'s come together, corresponding to these relations. Conversely, for any vanishing conditions
(of the usual translation and rotation invariant form) determining spaces of symmetric polynomials,
such an ideal $\cal I$ can be found.
So part at least of the structure of the CFT construction above is general [with $a(z)$ in place
of $\xi(z)$]. It is less clear whether the additional structure that the polynomials form a module
over the symmetric polynomials and over the positive part of the Virasoro algebra (which is automatic
in the MR construction), can be shown to be required, though as mentioned above, the first at least
seems to be necessary in order for many polynomials satisfying the vanishing conditions to exist.
In any case, finding a set of relations that generate such an ideal $\cal I$ corresponds to finding
special Hamiltonians with a minimal set of terms. Then the search for special Hamiltonians takes on
a truly algebraic form.

In recent work, it has been emphasized that the functions produced by the MR construction
for a given CFT can be viewed using the operator formalism for CFT. Then the operators $a(z)$ are
chiral vertex operators, acting in an auxiliary Hilbert space, which is that of the two-dimensional
Euclidean (i.e.\ one space, one time dimension) field theory, not the ``physical'' Hilbert space of
particles in two space dimensions. Then such a construction corresponds to a continuous version
\cite{drr2} of the popular idea of a matrix product state (MPS)---continuous, that is in the
sense that $z_i$'s are continuous variables, and a continuum, and in fact chiral, CFT is used.
In particular, the (infinite-dimensional) Hilbert space of the CFT corresponds to the auxiliary
space of the MPS construction. Closely related ideas are proving fruitful in on-going numerical
work on the trial states \cite{zm,eprb}.


\section{Analysis of a special Hamiltonian with a continuous parameter}
\label{scft}

In Ref.\ \cite{simon09}, it was shown that there is a simple form for the (one-parameter family of)
ground state wavefunctions in the MR construction applied to the ${\cal N}=1$ SCFT [where the field
$\psi(z)=G(z)$, the superconformal current], which satisfy
a corresponding family of ``vanishing conditions'' as three coordinates become equal. This result can
be interpreted as giving the ground state wavefunction for corresponding special Hamiltonians, whose
form can be explicitly determined. In this section we construct explicitly all the zero-energy
wavefunctions for these Hamiltonians and enumerate them. We emphasize we can view this as a direct
analysis of the spaces of functions determined by the given allowed behavior of the three-particle
functions that is one of this family of forms (or of zero-energy states for a corresponding special
Hamiltonian), with no use of CFT, other than that it motivated the parametrization in terms of the
central charge $c$. Though the analysis is somewhat similar to other examples (such as those in Refs.\
\cite{rr96,aks,read06}), the present example seems to be
the simplest in which there is a continuously-varying parameter in the functions.
The section is divided into three parts: the first gives results for the plane geometry, and the third for
the sphere with variable number of flux, which includes the plane results as a special limiting case.
The second part shows that the Hamiltonian lies on a phase boundary.

\subsection{Zero-energy states in the plane}

We have seen that for three particles, the symmetric polynomials of degree six in the internal or
relative coordinates span a two-dimensional
space. Then we may consider as a Hamiltonian the sum of projection operators onto all states of any group
of three particles with internal angular momentum (for the three) less than six, and onto one of the
two linearly-independent states of internal angular momentum six. Using Ref.\ \cite{simon09}, we
will parametrize the {\em allowed} behavior using first the four particle (non-symmetric) function
\be
\chi(z_1,z_2;z_3,z_4)=Az_{13}^3z_{24}^3z_{14}^3z_{23}^3
+z_{13}^4z_{24}^4z_{14}^2z_{23}^2,\ee
where $z_{ij}=z_i-z_j$ for all $i$, $j$, and $A=c/3 -1$ where $c$ and $A$ can be complex numbers. As
$z_1$, $z_2$, $z_3\to Z$, this becomes
\be
\chi\sim \chi_3(z_1,z_2;z_3)(Z-z_4)^6,
\ee
where
\bea
\chi_3(z_1,z_2;z_3)&=&\lim_{z_4\to\infty}z_4^{-6}\chi(z_1,z_2;z_3,z_4)\\
&=&Az_{13}^3z_{23}^3+z_{13}^4z_{23}^2.\eea
When symmetrized over $1$, $2$, $3$, this parametrizes the allowed (internal) behavior for three
particles in relative degree 6 (they must vanish in relative degree less than six). Apart
from overall normalization of the polynomial (which involves integration over all space, and depends
on the geometry used), this covers most of the space of quantum states in the two-dimensional space,
which are uniquely parametrized by a point on the sphere ${\bf CP}^1$. Namely, for finite complex
$c\neq\infty$ we cover all except one pole. By first dividing by $c$, we can cover all except the
opposite pole ($c=0$) of  ${\bf CP}^1$.

We will first study the behavior of symmetrized functions when two coordinates coincide. For the three
particle unsymmetrized function $\chi_3$, one finds that for $z_1=z_2$ it is given by $(A+1)z_{13}^6$.
This vanishes if $c=0$. Consequently, for $c=0$, the symmetrization of $\chi_3$ gives the $Q=2$
Laughlin function, $\prod_{i<j}(z_i-z_j)^2$ in the three coordinates, as the allowed behavior for
three particles in (internal) degree six, and similarly symmetrization of $\chi$ gives the Laughlin
function in four variables. Hence all the Laughlin $\nu=1/2$ ground and quasihole or edge functions are
allowed in this case, that is, functions in $N>2$ variables that vanish when any two coordinates are
equal, and so must be of the form $D_N f$ where $D_N=\prod_{i<j}(z_i-z_j)^2$ is the discriminant
(the square of the Laughlin-Jastrow factor) in $N$ variables, and $f$ is an arbitrary symmetric
polynomial. However, the functions are also allowed to be non-vanishing when two
coordinates coincide, but then they must vanish faster than degree six when a third coordinate
approaches those two. There are many additional functions that have this behavior.
(For $N=2$, the allowed functions are all of the symmetric polynomials in two variables.)
As this case is perhaps less interesting than $c\neq 0$, we postpone its analysis to Appendix
\ref{app:c0}, and proceed to $c\neq0$.

From here on, we restrict to $c\neq0$ (with $c=\infty$ handled as described above). Then
by referring to Ref.\ \cite{simon09}, we expect the ground state (the lowest degree polynomial of zero
energy) to have filling factor $\nu=1/3$. This is determined by the degree $N_\phi$ of the ground state
in each $z_i$, and a number called the shift, which is $6$, that is
\be
N_\phi=3N-6.
\ee
Then $\lim_{N\to\infty} N/N_\phi=\nu$. (The following exposition
will be somewhat terse, as the approach is similar to that in the earlier physics literature, in Refs.\
\cite{rr96,aks,read06}.) We consider the
map $C_1$ which sets two coordinates (without loss of generality, the last two) equal to $Z_1$. It maps
a polynomial from the space $\Lambda_N$ of symmetric polynomials in $N$ variables into the space
$\Lambda_{N-2}\otimes \Lambda_1$ of symmetric polynomials in $z_1$, \ldots, $z_{N-2}$ times a
symmetric polynomial in $Z_1$. This map has a kernel, that is, the space of functions it annihilates. The
functions in the kernel $\ker C_1$ are again of the form $D_N f$, for $f$ an arbitrary symmetric
polynomial (we write $D_N\Lambda_N$ for the space of such functions, and similarly in other examples).
More generally, consider the maps $C_m$ which set the last $2m$ coordinates equal in pairs,
to $Z_1$, \ldots, $Z_m$. That is, $C_m:f(z_1,\ldots,z_N)\to f(z_1,z_2,\ldots,
Z_1,Z_1,Z_2,Z_2,\ldots,Z_m,Z_m)$ and maps $\Lambda_N\to\Lambda_{N-2m}\otimes \Lambda_m$. Each of these
has a kernel in $\Lambda_N$, and $\ker C_{m-1}\subseteq \ker C_m$.

We wish to describe the space $\widetilde{I}_N$ (which is a module over the algebra $\Lambda$ of symmetric
polynomials) that is annihilated by our special Hamiltonian for arbitrary fixed $c$. That is,
functions in $\widetilde{I}_N$
vanish in a prescribed way in degree six as three particles come together, or faster. We define
$F_m=\ker C_m \cap \widetilde{I}_N$, for $m=0$, $1$, \ldots, $\lfloor N/2\rfloor$ (for $m=0$, $C_0={\rm
id}$, with kernel $\{0\}$). This immediately gives a filtration (sequence of inclusions of subspaces
into one another)
\be
F_0=\{0\}\subseteq F_1\subseteq F_2\subseteq\cdots F_{\lfloor N/2\rfloor}\subseteq
\widetilde{I}_N=F_{\lfloor N/2
\rfloor +1}.
\ee
If we consider $c=0$, then $F_m=\widetilde{I}_N$ for $m>0$. On the other hand, for $c\neq0$, all
the inclusions are strict, as can be seen because $\chi_3$ is nonvanishing in this limit
(or maybe better seen below).

The first kernel is $F_1=\widetilde{I}_N\cap D_N \Lambda_N$. But as we saw above, the
polynomials in $\widetilde{I}_N$ have three-particle relative behavior in degree six that is not
of the form $D_3$ unless $c=0$. For $c\neq0$ then, polynomials in the kernel $F_1$ have no term of degree
six (or lower) in the differences among any three coordinates, and so must be of the form $D_N$
times a symmetric polynomial that vanishes when any three coordinates are equal. But the symmetric
polynomials that vanish when any three coordinates are equal defines a module $\widetilde{I}_N^{\rm MR}$,
that is the zero-energy states for the MR state for bosons at $\nu=1$ \cite{milr,rr96}. That means we can
explicitly write down all functions in $F_1$.

This generalizes to give a description of $F_{m+1}$, modulo $F_m$, for all $m$,
including $m=\lfloor N/2\rfloor$ if we let $F_{\lfloor N/2\rfloor+1}=\widetilde{I}_N$. First, we
easily see from the allowed behavior in $\widetilde{I}_N$ that, for each $m$, the image of any function in
$\widetilde{I}_N$ (in particular, those in $F_{m+1}$) under $C_m$ is divisible by
\be
\prod_{i\leq N-2m}\prod_{k\leq m}(z_i-Z_k)^6\prod_{1\leq k<l\leq m}(Z_k-Z_l)^{12}.
\ee
Hence, now defining the induced map on the quotient space $C_m|_{F_{m+1}}/{\ker
C_m}:F_{m+1}/F_m\to\Lambda_{N-2m}\otimes \Lambda_m$, which is injective by definition, then we
find by a similar argument as the one just given (which was the case $m=0$) that its image is
contained in the space of polynomials of the form
\bea
\prod_{i<j\leq N-2m}(z_i-z_j)^2\cdot\prod_{i\leq N-2m}\prod_{k\leq m}(z_i-Z_k)^6\cdot&&
\nonumber\\
{}\times\prod_{1\leq k<l\leq m}(Z_k-Z_l)^{12}\cdot
\widetilde{I}_{N-2m}^{\rm MR}\otimes \Lambda_m&&\label{residues}
\eea
for $m=0$, $1$, \ldots, $\lfloor N/2\rfloor$.
(Here dots separate products of distinct scope, and we reuse indices in distinct products.) The functions
in these spaces correspond to the ``residues'' in earlier papers \cite{rr96,read06}. From this, it is
easy to count the number of such states, as the dimensions for the modules $\widetilde{I}_{N-2m}
^{\rm MR}$ are known. This is an
upper bound, as we have not yet shown that a state exists giving each linearly-independent residue
(that is, that the maps are surjective). It is clear that if they do exist, they are linearly independent.

To complete the argument first, we will construct one function that maps to each of the above residues.
First (following Ref.\ \cite{aks}), we introduce a labeling of the $N$ particles. The $N$ particles are
partitioned into $m_\alpha\geq0$ ($\alpha=1$, $2$, $3$) clusters of sizes $r_\alpha$, where
$r_\alpha=2$ ($\alpha=1$, $2$), $r_\alpha=1$ ($\alpha=3$), so $N=2(m_1+m_2)+m_3$. One particle is
assigned to each box in a Ferrers-Young diagram of at most two columns that corresponds to this
partition. Then the particle coordinates will be written as $z_{ij}^{(\alpha)}$, where $i$, $j$ label
rows and columns respectively (as for a matrix $i$ increases down the rows, while $j$ increases to the
right along the rows) of the rectangular block consisting of $m_\alpha$ rows of length $r_\alpha$.
That is, the first $m_1$ rows form the first block, the following $m_2$ the second, and the
remaining $m_3$ rows of length $1$ form the third block. Thus the rows of the diagram are ordered
with $(\alpha,i)$ above $(\alpha',i')$, written $(\alpha,i)>(\alpha',i')$, if either $\alpha<\alpha'$,
or $\alpha=\alpha'$, $i<i'$. We will deal with the MR part at the same time as the other parts.
It will turn out that $m_1=m$ in the above residues, while $m_2$, $m_3$ play a similar role for
the MR part. The assignment of particle coordinates to $z_{ij}^{(\alpha)}$ will be summed over
in the end, restoring the symmetrization.

The wavefunctions for given $m_\alpha$ and sets of integers $n_l^{(\alpha)}\geq 0$ ($l=1$, \ldots,
$m_\alpha$, and $\alpha=1$, $2$, $3$) are
\bea
&&{\cal S}_z\left\{\prod_{(1,i)>(1,i')}\chi(z_{i1}^{(1)},z_{i2}^{(1)};
z_{i'1}^{(1)},z_{i'2}^{(1)})\cdot\right.\\
&&{}\times\prod_{\alpha'=2,3}\prod_{(1,i)
>(\alpha',i')}\prod_{j=1}^{r_{\alpha'}}\chi_3(z_{i1}^{(1)},z_{i2}^{(1)}
; z_{i'j}^{(\alpha')})\cdot\non\\
&&{}\times\prod_{(1,m_1)>(\alpha,i)>(\alpha',i')}\prod_{j=1}^{r_\alpha}
(z_{i,j}^{(\alpha)}-z_{i',j}^{(\alpha')})
(z_{i,j+1}^{(\alpha)}-z_{i',j}^{(\alpha')})\cdot\non\\
&&{}\left.\times D_{N-2m_1}(\{z_{ij}^{(\alpha=2,3)}\})\cdot
\prod_{\alpha}\prod_{l=1}^{m_\alpha} e_l\left(\left\{\sum_{j=1}^{r_\alpha}z_{ij}^{(\alpha)}\right\}
\right)^{n_l^{(\alpha)}}\right\}
\label{scft_fns}
\non
\eea
Here $j+1$ is treated cyclically mod $r_\alpha$, so $z_{i,r_\alpha+1}^{(\alpha)}=z_{i,1}^{(\alpha)}$,
$D_{N-2m_1}(\{z_{ij}^{(\alpha=2,3)}\})$ is the discriminant in all the variables of types $\alpha=2$
and $3$, the elementary symmetric polynomials $e_l(\{\sum_{j=1}^{r_\alpha}z_{ij}^{(\alpha)}\})$ are
in the variables $\sum_{j=1}^{r_\alpha}z_{ij}^{(\alpha)}$, one for each row of type $\alpha$. To
see that these belong to $\widetilde{I}_N$, first notice that when two coordinates are equal, the
polynomial
under the symmetrizer ${\cal S}_z$ vanishes unless they both have $\alpha=1$ and $i$ is the same,
that is they are members of the same pair, which we term an ``unbroken pair'' when $\alpha=1$. When
three coordinates approach the same value, the function either tends to zero as ${\cal S}\chi_3$ in
these variables, which is the allowed form prescribed, or else vanishes faster. If coordinates are set
equal in pairs until there are $m$ pairs and doing so again would make the polynomial vanish, then all
these pairs are unbroken ($\alpha=1$ for each) and $m=m_1$. Further, the resulting function (consisting
of ``half-broken'' pairs with $\alpha=2$, and ``unpaired'' particles with $\alpha=3$) has the form
of (\ref{residues}), in which a polynomial in $I_{N-2m_1}^{\rm MR}$ can be identified using similar
results from e.g.\ Ref.\ \cite{aks}. This completes the argument, and as a whole it shows
that these functions form a complete and linearly-independent set. The ground (lowest degree) state,
which corresponds
to $m_1=N/2$, $m_2=m_3=0$, was found in Ref.\ \cite{simon09}, and has degree $N_\phi=3N-6$, as mentioned
above.


The number of states for each degree (angular momentum) can now be derived either from the explicit basis,
or from the residues, either for the whole thing or using results for $\widetilde{I}_{N-2m_1}^{\rm MR}$.
The degree of each function, omitting the symmetric polynomial factors, is
\bea
&&6m_1(m_1-1) +6m_1(2m_2+m_3)+2m_2(m_2-1) \non\\
&&{}+(2m_2+m_3)(2m_2+m_3-1)
+2m_2m_3 +m_3(m_3-1) \non\\
&&\quad{}=\frac{3}{2}N(N-2) +2m_2+m_3+\frac{1}{2}m_3^2
\eea
where we eliminated $m_1$ in favor of $N=2m_1+2m_2+m_3$. Then the character or generating function for
the dimensions of polynomials at each degree in $\widetilde{I}_N$ is
\bea
\lefteqn{{\rm ch}_q\, \widetilde{I}_N=}&&\\&&q^{\frac{3}{2}N(N-2)}\!\!\!\!\!\!\!\!\!\sum_{\stackrel{m_2,m_3
\geq0:2m_2+m_3\leq N,}{ (-1)^{m_3}=(-1)^N}}\frac{q^{2m_2+\frac{1}{2}m_3(m_3+2)}}{(q
)_{\frac{N-2m_2-m_3}{2}}(q)_{m_2}(q)_{m_3}},\nonumber
\label{ch_gen_sc}
\eea
where $(q)_m$ was defined in Eq.\ (\ref{q_m}).
This gives the complete counting of zero-energy states for our special Hamiltonian, for any finite
number of particles in the plane.

As a check on ${\rm ch}_q\, \widetilde{I}_N$, we can calculate the character of $\widetilde{I}_3$,
which should be all symmetric
polynomials, except for those of degrees 0, 2, 3, 4, 5, and one at 6 in the relative variables
(times center of mass factors).
The formula for ${\rm ch}\, \widetilde{I}_3$ does agree with this; it can also be written as ${\rm ch}\,
\widetilde{I}_3=q^6(1+q+q^2-q^4-q^5)/(q)_3$. Also ${\rm ch}_q\, \widetilde{I}_2=1/(q)_2$, as it should be.

We can take the limit $N\to\infty$ fixing attention on the zero-energy states of fixed degree relative
to the ground state simply by taking the limit of this formula with $m_2$, $m_3$ fixed. That is
\bea
&&\lim_{N\to\infty}q^{-\frac{3}{2}N(N-2)}{\rm ch}_q\, \widetilde{I}_N
\non\\
&&{}={}\sum_{m_2,m_3\geq0: (-1)^{m_3}=(-1)^N}\frac{q^{2m_2+\frac{1}{2}m_3(m_3+2)}}{(q
)_\infty(q)_{m_2}(q)_{m_3}}
\label{scft_n_infty}
\eea
The sum is still restricted in the parity of $m_3$, which depends on whether $N\to\infty$ through even
or odd values. This result has factored into separate factors for each $m_\alpha$. The factor
$1/(q)_\infty=\lim_{m\to\infty} 1/(q)_m$ is the limit of the character for the symmetric polynomials,
any one of which may be multiplied into any function in the space, and in FQHE often represent
edge excitations in the charge sector. The remainder of this expression will be interpreted in terms of
CFT later.

An extension of what we did here would be to consider the states with some quasiholes at the center of
the disk. For finite size these are merely special cases of the full set we obtained above. But when
taking $N\to\infty$ we can obtain different limits, and then we get the characters for other sectors of
the theory. In this way we can determine all the sectors without considering either the sphere or torus.

\subsection{Location at a phase boundary}

In the early stages of this analysis, we saw for $c\neq0$ that if we require the functions
to vanish when any two coordinates coincide, then there is no possible allowed behavior in relative
degree six (or less) for three particles. It followed that the allowed functions are of the form $D_N$
times a function in $\widetilde{I}_N^{\rm MR}$. These are the zero-energy functions associated with
the MR state at
$\nu=1/3$, and are independent of $c$. Hence, a special Hamiltonian with a two-body projection operator
(with positive coefficient) in addition to the six three-body ones (including the single $c$-dependent
one in degree six) already included produces for its zero-energy spaces the functions in
$D_N \widetilde{I}_N^{\rm MR}$. The fact that there is a {\em continuous family} of Hamiltonians
that produce
$\nu=1/3$ MR states for bosons does not seem to have been emphasized previously to our knowledge.

We will presume that the latter Hamiltonians generally have a gap in the bulk energy spectrum, and so lie
in the MR phase (for any $c\neq0$). But when the coefficient of the two-body term becomes zero,
the zero-energy spaces are much larger, and the energy spectrum of bulk excitations should become gapless
in the infinite-size limit, because for generic $c$ this system corresponds to amplitudes from a
non-unitary CFT. We conclude that, for any value $c\neq0$, there is a phase transition when
the coefficient of the two-body term is reduced to zero. The transition appears to have the same
behavior for all non-zero $c$, as the zero-energy states have the same multiplicities, however
their wavefunctions do depend on $c$. These states can be interpreted as containing varying numbers of
the ``unbroken pairs'' which are more tightly bound than the ``half-broken pairs'' present in the MR
ground state. When the coefficient of the two-body term becomes negative, the system presumably enters a
phase different from MR, the nature of which is not obvious. However, we may speculate that its ground
state consists of the more tightly bound pairs produced by the attractive two-body interaction,
something like a strong-pairing phase. The presence of the phase transition is similar to that in some
other examples discovered earlier \cite{rr96,green,rg}.

\subsection{Zero-energy states on the sphere}

It is simple to repeat the earlier analysis of the residues ($z_1$, $z_2\to Z_1$, etc) for $c\neq0$ for
the case of the sphere, in which the functions of $z$ should be the same degree for each particle $i=1$,
\ldots, $N$. For the residues, the functions take a Haldane-Laughlin--like form in the positions of
clusters $Z_i^{(\alpha)}$, $i=1$, \ldots, $m_\alpha$ for $\alpha=1$, $2$, $3$ (the $Z_i^{(3)}$ are
simply the coordinates of the unpaired particles) and quasiholes  $w_l^{(\alpha)}$, $l=1$, \ldots,
$n_\alpha$ for $\alpha=1$, $2$, $3$. Here the $w_l^{(\alpha)}$ appear in factors
$\prod_{i,l}(Z_i^{(\alpha)}-w_l^{(\alpha)})$ for each $\alpha$. When counting the degree of the
wavefunction in each particle variable $z_i$, one must realize that each $Z_i^{(\alpha)}$ for
$\alpha=1$, $2$ can come from either of two particles, so contributes only a half to the total
degree in $z_i$. Then one finds that the wavefunctions are of the same degree
\be
N_\phi=(6N+n)/2-6
\ee
(where $n\geq 0$ is even) in each $z_i$ regardless of type $\alpha$ provided
\bea
n_1&=&n,\\
n_2&=&n-4,\\
n_3&=&\frac{1}{2}n_2-m_3.
\eea
The last line follows from $n_1=2(m_3+n_3)+4$, and agrees with a corresponding relation in the MR state
in the formulation of Ref.\ \cite{read06}.

To write down the complete set of states such that one produces each residue, let us first note that
the factors $\chi$ can be used to couple a pair of type 1 to a pair of type 2, instead of two
factors $\chi_3$ as used above. There is no problem viewing these as functions on the sphere. The
factors $\chi_3(z_1,z_2;z_3)$ are still required to couple type 1 to type 3 (unpaired particles).
They appear problematic as the two terms are not the same degree in $z_1$ and $z_2$, and are of a
different degree in $z_3$. The first problem can be fixed by incorporating two quasihole factors
directly into $\chi_3$:
\bea
\lefteqn{\chi_3(z_1,z_2;z_3;w_1,w_2)=}\nonumber\\
&&A(z_1-z_3)^3(z_2-z_3)^3(z_1-w_1)(z_2-w_2)\\
&&{}+(z_1-z_3)^4(z_2-z_3)^2(z_2-w_1)(z_2-w_2).\nonumber
\eea
As $z_1$, $z_2\to Z$, this tends to the $\chi_3$ above for the plane, times $(Z-w_1)(Z-w_2)$. The
lack of symmetry between $w_1$ and $w_2$ is not important, because of the eventual symmetrization
over $z_i$'s. The relation $n_1=2(m_3+n_3)+4$ shows that there will always be more than enough
quasiholes of type $1$ to allow this to be done for all $\chi_3$ factors that appear. $\chi_3$
can be multiplied by additional factors $\prod_{l=1,2}(z_1-w_l)(z_2-w_l)$ to obtain the same
degree in $z_1$, $z_2$, as in $z_3$, fixing the second problem also.

The wavefunctions on the sphere are
\bea
&&{\cal S}_z\left\{\prod_{\alpha=1,2}\prod_{(1,i)>(\alpha,i')}
\chi(z_{i1}^{(1)},z_{i2}^{(1)};
z_{i'1}^{(\alpha)},z_{i'2}^{(\alpha)})\cdot\right.\\
&&{}\times\prod_{(1,i)
>(3,i')}\chi_3(z_{i1}^{(1)},z_{i2}^{(1)}
; z_{i'1}^{(3)};w_{2i'-1}^{(1)},w_{2i'}^{(1)})\cdot\non\\
&&{}\times\prod_{(1,m_1)>(\alpha,i)>(\alpha',i')}\prod_{j=1}^{r_\alpha}
(z_{i,j}^{(\alpha)}-z_{i',j}^{(\alpha')})
(z_{i,j+1}^{(\alpha)}-z_{i',j}^{(\alpha')})\cdot\non\\
&&{}\left.\times D_{N-2m_1}(\{z_{ij}^{(\alpha=2,3)}\})\cdot\prod_\alpha\prod_{i,j}
{\prod_l}^{(\alpha,j)} (z^{(\alpha)}_{ij}-w^{(\alpha)}_l)
\right\}.
\non
\eea
The notation is identical to that for the plane, except for the introduction of the $\chi_3$ above,
and the final product of quasihole factors, in which ${\prod_l}^{(\alpha,j)}$ appears, where the
range of $l$ depends on $\alpha$, $j$:
\be
\begin{array}{rcllcl}
l&=&2m_3+1,&\ldots,& 2m_3+n_3+2&(\alpha=1,j=1), \\
l&=&2m_3+n_3+3,&\ldots, &n_1&(\alpha=1,j=2),\\
l&=&1,&\ldots,&n_2/2&(\alpha=2,j=1),\\
l&=&n_2/2+1,&\ldots,&n_2&(\alpha=2,j=2),\\
l&=&1,&\ldots,&n_3&(\alpha=3,j=1).
\end{array}
\ee
Each of these functions produces just one of the residues when particles coincide in pairs. As the
residues are symmetric under permutations of $w_l^{(\alpha)}$ (for fixed $\alpha$), it is
permissible to symmetrize over the $w_l^{(\alpha)}$ of each $\alpha$ in the full wavefunctions.

Now we calculate the character, which here is the group-theoretical character for the representation of
the rotation group of the sphere acting on the zero-energy states, or in other words the generating
function that counts the number of states of each $L_z$. We use the $q$-binomial coefficient in the form
\be%
{m_\alpha+n_\alpha\choose
n_\alpha}_{\!q}=\frac{(m_\alpha+n_\alpha)_q!}{(m_\alpha)_q!
(n_\alpha)_q!},\ee%
where the $q$-factorial $(n)_q!=[n]_q[n-1]_q\cdots[1]_q$, and the
$q$-deformed integers are
\bea
[n]_q&=&q^{-(n-1)/2}+q^{-(n-3)/2}+\cdots+q^{(n-1)/2}\non\\
&=&\frac{q^{-n/2}-q^{n/2}}{q^{-1/2}-q^{1/2}}.
\eea
(This differs from what was used in Ref.\ \cite{read06}, but is more convenient for rotationally-invariant
situations.) For $q\to 1$, the $q$-binomial reduces to the usual binomial coefficient. Then by counting
the symmetric polynomials in the $w_l^{(\alpha)}$ that occur in the residues, the generating function is
\be
{\rm tr}\,q^{-L_z}=\sum_{m_1,m_2,m_3:2m_1+2m_2+m_3=N}\prod_{\alpha=1,2,3}
{m_\alpha+n_\alpha\choose
n_\alpha}_{\!q}.
\ee

The results for counting states in the infinite plane can be recovered by letting $n\to\infty$, as
follows. In the plane, by stereographic projection from the sphere, $-L_z$ corresponds to angular
momentum in the plane (total degree), but we must shift $-L_z$ by adding $NN_\phi/2=N(6N+n)/4-3N$.
Combined with the highest $L_z$ (which occurs for $N$ even, $m_1=N/2$, $m_2=m_3=0$), which is $L_z=Nn/4$,
we recover
the degree of the ground state in the plane, $\frac{3}{2}N(N-2)$ (independent of $n$).
After extracting this factor, the $q$-binomials take on the definition used in Ref.\ \cite{read06},
in which
$[m]_q\to(m)_q=(1-q^m)/(1-q)$, and this is convenient for taking limits such as $n\to\infty$ at fixed $N$.
In this way the generating function (character) in the plane, Eq.\  (\ref{ch_gen_sc}), can be recovered.

Returning to the sphere, we can view the $q$-binomial in $m_1$, $n_1$ as describing the
``positional'' degeneracy of what would be the quasiholes if the bulk energy spectrum were gapped,
as in a topological phase: there are $n=n_1$ quasiholes of charge $1/6$. The other factors describe the
degeneracy for fixed positions. This depends on $m_2$, $m_3$, while $m_1=(2m_2+m+3)/2$ is the number of
unbroken pairs. The total degeneracy for fixed positions is
\be
\sum_{m_2,m_3} {m_2+n_2\choose n_2}{m_3+n_3\choose n_3}
\ee
where the sum is over $m_2$ and $m_3$ such that $N-2m_2-m_3$ is even and $\geq0$, and $n_2$, $n_3$ are
determined by $n$ and $m_3$ as above. In a topological phase, this would approach a finite limit as $N$ (or
$N_\phi$) $\to\infty$ with $n$ fixed. In particular,
\be
{m_3+n_3\choose n_3}={n_2/2\choose m_3}
\ee
corresponds to $m_3$ fermions that can occupy $n_2/2$ states. This factor is never large, and
$m_3$ must be less than or equal to $n_2/2=n/2-2$. But the binomial coefficient in $m_2$, $n_2$
corresponds to $m_2$ bosons in $n_2+1$ orbitals, and as $N\to\infty$ with $n_2>0$ there is no upper limit
on $m_2$. That is, the number of half-broken pairs can be arbitrarily large as $N\to \infty$. Hence the
total degeneracy for fixed positions of the quasiholes becomes arbitrarily large, similar to the
permanent and Haffnian cases \cite{rr96,green}. This behavior cannot occur in a gapped (topological)
phase of matter.


\section{Amplitudes and modules}
\label{sec:amps}

\subsection{CFT---operator methods}
\label{sec:cft}

In this section, we first set up definitions and basic general arguments that will be used in the
following. The starting point is the general form of the MR construction of trial wavefunctions, eq.\
(\ref{ground}) for the ground state. Then we describe our first main result.

In this paper, the main examples are paired states for which the operator product expansion (ope)
of $\psi$ reads
\be
\psi(z)\psi(0)\sim \frac{1}{z^{2h_\psi}}+\ldots
\ee
as $z\to0$,
where $h_\psi$ is the conformal weight of $\psi$, and $\ldots$ denotes lower-order terms (not only
non-singular terms). In our main examples, either (i) $\psi$ is the field at the corner of the
minimal block in a BPZ minimal
model $M(p,p')$, that is $\psi=\phi_{(1,p)}$ or $\phi_{(p',1)}$ in the $p$, $p'$ minimal model (we use
the conventions of Ref.\ \cite{dfms}), or (ii) $\psi$ is the superconformal current $G(z)$ in an ${\cal
N}=1$ superconformal field theory. The first of these includes, but is more general than, the
$\phi_{(2,1)}$ field used by FJM. (For the Laughlin states, the analysis is
similar but easier.) We will, however, extend the general results to the case of generalized parafermionic
CFTs \cite{zamo}, or clustered FQHE states, of order $k=2$, $3$, $4$ \ldots; in these CFTs, when the ope 
of $\psi$'s
is iterated, $k$ $\psi$'s are needed in order to obtain the identity operator on the right hand side, as
for the $k=2$ case above (the case $k=1$, where $\psi=I$, the identity operator, gives the Laughlin
states only). In all cases, the right hand side of the ope of $\psi$ with itself is assumed to contain
only operators of the same scaling dimension as the operator in leading term, plus integers, and
similarly for the iterated opes; $\psi$ should be a ``simple current'' (these conditions ensure that,
for suitable values of $h$ and $h_\psi$,
the wavefunctions we obtain are indeed just symmetric polynomials). The assumed behavior of the
iterated ope's of $\psi$ leads to restrictions on the value of $h_\psi$ (we will not need the details).
For paired ($k=2$) cases, single-valuedness and symmetry implies that $\nu^{-1}-2h_\psi$ must be a
non-negative even integer. In particular, one can consider $\nu=1/(2h_\psi)$ so the wavefunctions
are non-vanishing when two coordinates coincide. For general $k$, we require that
the conformal weight $h=\nu^{-1}/2+h_\psi$ of $a$ be half a positive integer so that $a$ and
its adjoint generate a chiral algebra, or vertex operator algebra (VOA) \cite{dfms,kac_book}, as we will
discuss below.
In some cases $2h$ is odd, even though the wavefunctions that are produced describe bosons---the
chiral algebra generated by $a$ and $a^\ast$ violates the spin-statistics theorem (this occurs only
when $\psi$ is part of a non-unitary theory, such as the case of the Gaffnian for bosons). From
single-valuedness of the ope of $a(z')$ with $e^{\pm ik\varphi(z)/\sqrt{\nu}}$ (which appears
in the iterated ope of $k$ $a(z)$s), we find that $k/\nu$
must be an integer. We can also obtain antisymmetric wavefunctions for fermions in a similar way,
either by relaxing these conditions (leading to chiral superalgebras, again either obeying or not
obeying the spin-statistics relation), or simply by multiplying the wavefunctions for bosons by the
Vandermonde determinant $\prod_{i<j}(z_i-z_j)$.

We have the charge current operator $j(z)=i\partial \varphi$, with operator product
\be
j(z)e^{in\varphi(0)/\sqrt{\nu}}\sim \frac{n/\sqrt{\nu}}{z} e^{in\varphi(0)/\sqrt{\nu}}.
\ee
Thus $j$ measures the charge, up to the factor $\sqrt{\nu}$. Being a CFT, the theory has a stress tensor
\be
T(z)=T_\varphi(z)+T_\psi(z),
\ee
which is a sum of the stress tensors $T_\varphi$ and $T_\psi$ of the theories to which $\varphi$ and
$\psi$ belong. We have
$T_\varphi(z)=-\frac{1}{2}(\partial\varphi)^2$. Finally, we have a vacuum $|0\rangle$ of the combined CFT,
which is the tensor product of the vacuua of the $\varphi$ and $\psi$ theories,
$|0\rangle=|0_\varphi\rangle\otimes |0_\psi\rangle$. We emphasize that in this section, all states
and operators are those of the two- ($1+1$-) dimensional CFT, not the $2+1$ dimensional many-particle
system.

The mode expansion of $a$ can be defined by
\be
a(z)=\sum_{n+h\in {\bf Z}} a_n z^{-n-h}
\ee
as $z\to 0$ (acting to the right). The modes can therefore be extracted as
\be
a_n=\frac{1}{2\pi i}\oint dz \,z^{n-1+h}a(z).
\ee
The integral is over a circle of radius $r$. As there is no singularity at the
origin if no non-trivial operator is inserted there, we must have
\be
a_n|0\rangle=0
\ee
for $n> -h$, and we can identify the state $a(0)|0\rangle=|a\rangle=a_{-h}|0\rangle$.  All products
of $z$-dependent operators are radially-ordered, unless otherwise stated. To obtain the product of
modes in any given order, one uses the corresponding ordering of the radii of the circles in the
integral defining each mode.

The mode expansion takes the same form for any conformal (Virasoro primary) field, with $h$ replaced by
the conformal weight, and the mode indices $n$ must obey $n+h\in {\bf Z}$. In particular, the modes
$j_n$ of the weight-$1$ U$(1)$ current $j(z)$, and $L_n$ of the weight-$2$ stress tensor $T(z)$,
obey the usual relations, in particular the Virasoro algebra relations for the modes
$L_n=L_{n\varphi}+L_{n\psi}$ of $T$ in radial quantization \cite{dfms}. Also, the modes obey relations
with $a(z)$,
\bea
[j_n,a(z)]&=&\frac{1}{\sqrt{\nu}}z^na(z),
\label{jcommrel}\\
{}[L_n,a(z)]&=&h(n+1)z^na(z)+z^{n+1}\frac{\partial a(z)}{\partial z}.
\label{Lcommrel}
\eea
Finally, we also have
\bea
j_n|0\rangle&=&0\hbox{ ($n\geq0$)},\label{jvac}\\
L_n|0\rangle&=&0\hbox{ ($n\geq -1$)},\label{Lvac}
\eea
and we refer to the sets of modes in the last relations as the positive half of the current and
Virasoro algebras, respectively.

For our main examples, in which $k=2$, it is clear that the conjugate field of $a$ (which creates
an ``antiparticle'' from the vacuum, if $a$ creates a ``particle'') is
\be
a^\ast(z)=e^{-i\varphi(z)/\sqrt{\nu}}\psi(z).
\ee
This construction works because $\psi$ is its own conjugate. $a^\ast$ has the same conformal weight
$h$ as $a$. Clearly $a^\ast$ can be obtained by the local operation of ordinary complex conjugation,
with $\varphi(z)$, $\psi(z)$ treated as {\em invariant} (self-conjugate).

We now assert that this operation can be viewed as the radial adjoint (here we differ slightly from
some standard treatments, such as Ref.\ \cite{dfms}):
\be
a^\ast(z)=a(z)^\dagger.
\ee
For any conformal field, the radial adjoint is defined using radial quantization, and the radial
coordinate plays the role of imaginary time (because of the Euclidean signature on the
two-dimensional spacetime; note also that the definition of the adjoint operation depends on a
choice of origin). That is, we first map conformally to the cylinder, with the axis of the
cylinder as imaginary time; the map is $z\mapsto -i\ln z$. We then apply the adjoint operation,
which strictly speaking is only the usual operator adjoint at $|z|=1$ (imaginary time equal to zero);
for non-zero time, it involves reversing the sign of imaginary time (in addition to complex conjugation),
as in the imaginary-time formalism in many-body theory \cite{fw}. That is, $-i\ln z$ is invariant in
this step. Finally, we apply the inverse conformal map back to the plane. The conformal map to the
cylinder means multiplying by a factor $z^h$, and by the inverse for the inverse map. Thus in this
point of view, $z^\dagger=z^{-1}$, except in the $z^{-h}$ factors which are invariant. Putting all
these steps together, we have
\be
a(z)^\dagger=\sum_n a_n^\dagger z^{n-h}.
\ee
In modes, using the standard form of mode expansion for $a^\ast$ or $a^\dagger$, this then implies
\be
(a^\dagger)_n=(a_{-n})^\dagger.
\ee
(This last formula is standard \cite{bpz}.) The latter formulas apply to all conformal fields, for
example $\psi(z)$, and also to $j(z)$, $T(z)$ which are both self-conjugate, so that
$j_{-n}=j_n^\dagger$, and so on.

Usually adjoints are defined relative to an inner product, but in CFT we frequently do the reverse.
We take the preceding formulas for the adjoint operators, together with the inner product of the
vacuum $\langle0|0\rangle=1$. Inner products of states defined as any finite collection of modes
of the fields acting on $|0\rangle$ can now be defined using the preceding formulas, and using
the commutation relations to simplify expectation values of the modes. This determines the inner
product on the space of states, which is Hermitian but not necessarily positive definite or
non-degenerate.

We define
\be
|0_N\rangle=e^{iN\varphi(0)/\sqrt{\nu}}|0\rangle,
\ee
or in an obvious notation $|0_N\rangle=|0_{\varphi N}\rangle\otimes|0_\psi\rangle$.
Also we can define an out vacuum of charge $-N$ by
\be
\langle0_N|=\lim_{Z\to\infty}Z^{N^2/\nu}\langle 0|e^{-iN\varphi(Z)/\sqrt{\nu}}.
\ee
[We can do similarly for out states of other conformal fields, using twice their conformal weight
in place of $N^2/\nu$, thus
$\langle a|=\lim_{z\to\infty}z^{2h}\langle 0|a^\ast(z)$.]
Then we have
\be
\langle 0_N|0_{N'}\rangle=\delta_{N,N'}.
\ee

We can now define the chiral algebra, or vertex operator algebra (VOA) \cite{kac_book}, of our theory
as the algebra generated by the modes of fields generated by repeated ope's of $a(z)$ and $a^\ast(z)$.
(We do not mean to imply that a chiral algebra and a VOA are the same concepts when defined rigorously,
however for our purposes greater precision will not be required.) We will argue that
it includes the current operator $j(z)$ and both stress tensor operators, $T_\varphi(a)$ and $T_\psi(z)$.

For the $\varphi$ theory, we define the space of states ${\cal F}_N$ ($N\in{\bf Z}$), to be the
Fock spaces generated from $|0_{\varphi N}\rangle$ by the action of the modes $j_n$ ($n<0$). For
the $\psi$ theory, we define the space $M_0$ ($M_1$) to be the space of states generated by the
action of an even (odd) number of $\psi$'s on the vacuum $|0_\psi\rangle$. In the Virasoro minimal
model case, $M_0$ is the vacuum $(1,1)$ irreducible Virasoro module, and $M_1$ is the irreducible
Virasoro module generated by $|\psi\rangle$. In the SCFT case, we will discuss the structure
of the spaces later. Then, as in Ref.\ \cite{fjm}, the space $V$ of all states spanned by $a$ and
$a^\ast$ acting on $|0\rangle$ is $V=\bigoplus_NV_N$, where
\be
V_N=\left\{\begin{array}{l}{\cal F}_N\otimes M_0\hbox{ ($N$ even),}\\
{\cal F}_N\otimes M_1 \hbox{ ($N$ odd).}
\label{V_struc}
\end{array}\right.
\ee
The inclusion $V\subseteq \bigoplus_N V_N$ is clear, but the equality requires proof. First, we notice
that $V$ is a module over (i.e.\ a representation of) the VOA. By the state-operator correspondence
(or the theory of VOAs) \cite{bpz,dfms,kac_book}, $V$ ``has the structure of'' the VOA, in particular
there is one
(linearly-independent) state for each non-zero, linearly-independent operator in the VOA, and
the Virasoro level and the charge of the states and operators agree.

It is not too difficult to see that $V$ as defined here is indeed generated by $a(z)$ and $a(z)^\ast$.
Essentially, from the ope's one can extract $j(z)$, and hence separate out the spaces ${\cal F}_N$.
The generalization from order $k=2$ to $k>2$ is straightforward, involving $k$ modules $M_0$,
\ldots, $M_{k-1}$ in place of the two above. A more detailed proof for all cases is given
in Appendix \ref{app:lim}.

The mode operators $a_n$ commute, because the operator product $a(z_1)a(z_2)$ is non-singular.
The operators $a$ generate a commutative subalgebra $U$ (the principal subalgebra) of $V$, and the
subspace spanned by these operators acting on $|0_N\rangle$ will be denoted $V^N$:
\be
V^N=U|0_N\rangle
\ee
Then we have a chain of embeddings (injective maps)
\be
\cdots\longrightarrow V^4 \longrightarrow V^2\longrightarrow V^0\longrightarrow V^{-2}\cdots,
\ee
all of which are subspaces of $V$.
The embeddings arise because $V^{N-2}$ includes $a(z_1)a(z_2)|0_{N-2}\rangle$, and the leading part
as $z_1$, $z_2\to0$ gives $|0_N\rangle$, and hence all of $V^N$ appears inside $V^{N-2}$.
We call $W=V^0$ the principal subspace (it corresponds to the principal subalgebra). There is
also \cite{fjm} an operator that we will call $\cal Q$ that shifts the charge by $2$ (it is part of
the operator construction of $e^{2i\varphi(z)/\sqrt{\nu}}$), and acts as
\be
{\cal Q}a_n{\cal Q}^{-1}=a_{n+ 2/\nu}
\ee
on the modes. The map ${\cal Q}:V^N\to V^{N+2}$ is an isomorphism. ${\cal Q}$ will be used later.
The appearances of the number $2$ in these statements should be replaced by $k$ in general.

It was shown in Ref.\ \cite{fs} and FJM, for the particular set of models they consider, that $V$
is isomorphic to the
direct (or inductive) limit $\lim_{\longrightarrow} V^{-N}$ of the above chain of embeddings, defined by
\be
\lim_{\longrightarrow} V^{-N}\equiv\lim_{N\to\infty}V^{-N}\equiv \bigcup_N V^{-N}.
\ee
Their proof uses the irreducibility of $V_N$ ($V_N$ is the subspace of $V$ of charge $N$) under the
action of the algebra generated by the current $j(z)$ and the stress tensor $T(z)$. This holds in
our minimal model constructions, of which the models in FJM are special cases. A more general version,
for all $k=2$ constructions, can be proved using the algebra generated by $\psi$ itself, as can the
generalization to all $k>0$. That is, we have
\newline
{\bf Theorem 1}: In notation as above, for the order $k$ case,
\be
V \cong \lim_{\longrightarrow} V^{-N}.
\ee
Equivalently, $V$ is generated as a VOA by $a(z)$ and
\be
e^{-ik\varphi(z)/\sqrt{\nu}}.
\ee
(The latter is needed to reach the negatively charged vacuua $|0_{-N}\rangle$.)
We again defer the detailed proof, which is not difficult, to Appendix \ref{app:lim}. This is the
first main result of this Section. (For the Laughlin or $k=1$ case, the
Theorem is trivial, as $V$ is generated by $a(z)$ and $a(z)^\ast=e^{-i\varphi(z)/\sqrt{\nu}}$,
by definition.) We comment that from Theorem 1 there also follows a similar result for modules
over the VOA, other than the vacuum module $V$ itself (viewed here as a space of states). The same 
two $z$-dependent operators generate these sectors of states also, starting from a state 
$|\tau\rangle=\tau(0)|0\rangle$ that corresponds to a primary field $\tau(z)$.
(We are grateful to a referee for this remark, which is not used elsewhere in this paper.) 
For suitable $\tau$,
the analog of the principal subspace corresponds in the MR construction, or under the duality discussed 
in the next Subsection, to wavefunctions with a quasihole at the origin (see e.g. Ref.\ \cite{read09}).

The principal subspace $W$ contains vectors constructed from the commuting operators $a(z)$. $W$ is a
module (representation) of the principal subalgebra (i.e.\ itself, under the state-operator
correspondence).  From the relations (\ref{jcommrel}), (\ref{Lcommrel}), we have commutation relations
for the modes of $a(z)$,
\bea
[j_n,a_m]&=&\frac{1}{\sqrt{\nu}}a_{m+n}
\label{jLacommrel1}\\
{}[L_n,a_m]&=&[(h-1)n-m]a_{m+n},
\label{jLacommrel2}
\eea
and see also eqns.\ (\ref{jvac}), (\ref{Lvac}). $W$ is thus also a module over these positive halves
of the current and Virasoro algebras. This positive half of the Virasoro algebra includes the
generators $L_0$, $L_{\pm 1}$ of the Mobius group, but the central charge never shows up. Note that
$L_{-1}$ generates translations.

We can describe the structure of $W$ by considering an algebra $\cal A$ spanned by products of
the commuting indeterminates $\xi_{-n}$, $n=0$, $1$, \ldots (which correspond to $a_{-n-h}$ by dropping
the shift by $h$). The operators $a$ in general obey further relations, and these relations generate
an ideal ${\cal I}\subset {\cal A}$. Then we have the isomorphism
\be
W\cong {\cal A}/{\cal I}.
\ee
We emphasize that the ideal ${\cal I}$ of relations in $\cal A$
can be thought of as generated by {\em local} relations among the $a$'s, because of locality
in the CFT. Examples of such relations are $a(z)^2=0$ in the Laughlin $\nu=1/2$ example, and
$a(z)^3=0$ in the MR $\nu=1$ example, in which $\psi$ is a Majorana fermion field. The relations
must be independent of position $z$ (translation invariance), respect the degree $d=\sum n_i$ of
the products of $\xi_{-n}$s (rotation invariance), and respect the charge $N$ (charge conservation).
Each such subspace ${\cal A}_{d,N}$ of degree $d$ and charge $N$ is finite dimensional. Any such
expression written as modes of $a(z)$ acting on $|0\rangle$ can be viewed as built from $a$'s
and derivatives of $a$'s acting at $z=0$. The translation invariance
of the VOA allows such a relation to be moved to any point $z$.  Moreover, there is a natural action of
the (positive) current and Virasoro modes on $\cal A$, determined by the same relations
(\ref{jLacommrel1}), (\ref{jLacommrel2}). The subspaces ${\cal I}_N$ (of vectors with charge $N$)
must be mapped into themselves by these modes.

\subsection{Dual spaces of polynomials}
\label{sec:dual}

We now turn to the ``functional model'' of $W$ \cite{fs,fjm} which corresponds to wavefunctions
for bosons in the QH situation.
We recall that the dual ${\cal B}^\ast$ of a vector space ${\cal B}$ is the space of linear maps
from $\cal B$ to the complex numbers, and the action of a map $d\in {\cal B}^\ast$ on a vector
$b\in{\cal B}$ can be written as the ``dual pairing'' $\langle d,b\rangle$ (no complex
conjugation is involved).
The dual vector space of ${\cal A}$ can be represented by symmetric polynomials; we write
$\Lambda_N$ for the symmetric polynomials in $N$ variables (which are here usually written $z_1$,
\ldots, $z_N$), and $\Lambda_{d,N}$ for the subspace of polynomials of degree $d$. We have the natural
dual pairing \cite{fjm}
\bea
\lefteqn{\langle f(z_1,\ldots,z_N),\xi_{-n_1}\cdots\xi_{-n_N}\rangle=}&&\\
\,&&{\rm Res}_{z_1=\cdots z_N=0}f(z_1,\ldots,z_N)z_1^{-n_1-1}\cdots z_N^{-n_N-1},\non
\eea
where $f$ is a symmetric polynomial in $\Lambda_N$, and the residue is the usual one of complex
analysis, applied to all $z_i$. That is, we simply pick out the part of $f$ (expanded in the
``monomial basis'' of symmetrized products of powers of $z_i$) in which the degrees in the $N$
distinct variables are (in any order) $n_1$, \ldots, $n_N$. The dual space $W_N^\ast$ of $W_N$ (the
subspace of $W$ spanned by $N$ $a$'s acting on $|0\rangle$) is then the subspace $I_N$ of $\Lambda_N$
of polynomials that annihilate the module ${\cal I}_N$ with respect to the above pairing,
\be
I_N=\{f\in\Lambda_N|\langle f,\xi\rangle=0 \;\forall \xi\in{\cal I}_N\}.
\ee
This means that $I_N$ can be viewed as the space of symmetric polynomials that satisfy certain
``vanishing conditions'' on their behavior as several $z_i$'s come to the came value; these conditions
are derived from the local relations in $\cal I$.

Equivalently, we may think of $I_N$ as the space of all symmetric polynomials constructed as
\be
f_v(z_1,\ldots,z_N)=\langle v,a(z_1)\cdots a(z_N)|0\rangle\rangle
\ee
where $v$ is an element of the dual space $V^\ast$ of $V$, and $\langle\;,\;\rangle$ is the dual
pairing with $V$. Clearly such functions inherit vanishing properties from the relations satisfied
by the $a(z)$s. This form is very close to what we wish to use for wavefunctions, but it will be
important later to realize that the use of the {\em dual} space $V^\ast$ here is not always the
same as using the space of {\em out} states. The out states can be viewed as elements of the dual
space, which act on the space $V$ by the natural (Dirac) pairing. Thus the out space can be mapped
into the dual space; however this map may not be injective. We will now discuss this.

If the inner product (Hermitian sesquilinear form) on $V$ is non-degenerate, then we can identify the
dual space $V^\ast$ with $V$ using the inner product (more precisely, we have a conjugate-linear map
from one to the other). We recall that an inner product is degenerate (or singular) if there are
vectors that have vanishing inner product with all other vectors; such vectors are called
{\em singular} (or sometimes, miscalled null) vectors. The singular vectors form a subspace, and
the space modulo the subspace of singular vectors (formed by ``setting the singular vectors to
zero'') possesses a non-degenerate inner product.

Instead of using the dual space $V^\ast$ to obtain the polynomials, we can use the inner product and
obtain polynomials
\be
f_w(z_1,\ldots, z_N)=\langle w|a(z_1)\cdots a(z_N)|0\rangle,
\ee
where $w\in V$. We denote by $J_N$ the space of such polynomials for $N$ $a$'s. Clearly, $J_N\subseteq
I_N$. If the inner product is non-degenerate, then $J_N=I_N$. But if the inner product is degenerate,
then $J_N$ may not equal $I_N$, as we have already noticed above. In finite dimensional vector spaces,
the space and its dual have the same dimension. The space $I_N$ in each total degree $d$ (we write
these as $I_{d,N}$, and similarly for $J_{d,N}$) has the same dimension as $W_N$ at $L_0$
eigenvalue $d+Nh$ (which we write as $W_{d,N}$). But $J_{d,N}$ has the same dimension as the quotient
space, $W_{d,N}$ modulo singular vectors lying inside it. $J_N$ can be identified as the dual of
that quotient space.

The spaces of polynomials $I_N$ and $J_N$ at which we have arrived have further general properties
that reflect general aspects of our construction.
The right action of the current modes $j_n$ ($n\geq0$) and the Virasoro modes $L_n$ ($n\geq-1$) on
$v\in V^\ast$ give \cite{fjm}
\bea
f_{v\cdot j_n}(z_1,\ldots,z_N)&=&\frac{1}{\sqrt{\nu}}
s_nf_v(z_1,\ldots,z_N)\\
f_{v\cdot L_n}(z_1,\ldots,z_N)&=&h(n+1)s_nf_v(z_1,\ldots,z_N)
\non\\
&&{}+\ell_n(f_\Psi)(z_1,\ldots,z_N)
\eea
where for $n\geq0$
\be
s_n=\sum_{i=1}^N z_i^n
\ee
are the sums of powers ($s_0=N$), and for $n\geq -1$
\be
\ell_n=\sum_{i=1}^N z_i^{n+1}\frac{\partial}{\partial z_i}.
\ee
These follow from eqns.\ (\ref{jcommrel}), (\ref{Lcommrel}), (\ref{jvac}), and (\ref{Lvac}). The sums
of powers $s_n$ (with exponents $0<n\leq N$) are another set of generators for the symmetric
polynomials $\Lambda_N$, and so the space $I_N$ is mapped into itself by multiplication
by symmetric polynomials --- it is a module over the algebra of symmetric polynomials. Similarly,
$I_N$ is a module for the
positive half (actually $n\geq -1$) of the Virasoro algebra, acting by the $\ell_n$'s. That is,
$s_nI_N\subseteq I_N$ and $\ell_nI_N\subseteq I_N$. Both sets of operators raise the degree of the
polynomial (for $n>0$), and usually one might expect them to be associated with the raising (negative)
parts of the algebras; the apparent reversal of sign is due to the dual relationships in use here.
Identical arguments apply if the operators are applied (from the right) to the out-state vector
$\langle w|$ to obtain an action on the functions in $J_N$. If we think of the adjoints $j_{-n}$
and $L_{-n}$ acting on the vector $|w\rangle$, then these are indeed raising parts. For another view 
of the spaces of amplitudes of certain CFTs, which includes further differential operators, see also 
Ref.\ \cite{estienne2}.

It is useful to realize that, in terms of the interpretation of symmetric polynomials as wavefunctions
in the LLL, $\ell_{-1}$ is a derivative with respect to the center of mass variable, while $j_1$ acts
as multiplication by ($N$ times) the center of mass. These obey the Heisenberg relation
\be
[\ell_{-1},j_1]=-N,
\ee
corresponding to the projective representation of the translation group of the plane.

Turning again to the question of singular vectors,
for a degenerate inner product in the context of a CFT or VOA, the subspace of $V$ (or of any
other module obeying the adjointness conditions) consisting of all the singular vectors must form
a submodule of the VOA. Conversely, in an indecomposable module of the VOA (i.e.\ one that is not a
direct sum of modules), a submodule must consist entirely of singular vectors. Finally, the quotient of
$V$ by the subspace of singular vectors yields a non-degenerate VOA on which the inner product
is non-degenerate. We note that a non-degenerate inner product is a necessary condition both for a
rational and for a unitary CFT (but not sufficient for either).

In some situations, examples of which we will study in this paper, a VOA is first constructed that
depends on a parameter, such as the Virasoro central charge $c$. For generic values of $c$, the space
$V$ may be non-degenerate. But for special ``rational'' values of the parameter, singular vectors
may appear \cite{bpz,dfms}, and then they have to form a submodule of the VOA inside $V$. The central point
is that this subspace may intersect the principal subspace $W$. In terms of the algebra $\cal A$,
there is then a larger ideal ${\cal I}'$ (over $\cal A$) that corresponds to the singular vectors.
That is ${\cal I}\subset {\cal I}'\subseteq {\cal A}$. Then ${\cal I}'/{\cal I}\subset
{\cal A}/{\cal I}\cong W$ is the singular subspace of $W$. Then, as we know, $J_N$ is the
dual of the quotient of these spaces,
\be
J_N\cong \left(W_N/({\cal I}_N'/{\cal I}_N)\right)^\ast,
\ee
or $J_N\cong ({\cal A}_N/{\cal I}_N')^\ast$. In other words, the smaller space $J_N$ is dual to a space
that is modded out by additional (independent) relations among the $a$'s, described by ${\cal I}'$,
rather than ${\cal I}$. That is, the subspace $J_N\subset I_N$ is defined as polynomials that
obey additional (independent) ``vanishing conditions'' as several $z_i$'s approach the same point,
compared with those defining $I_N$ if $J_N\neq I_N$. In particular examples, as a continuous parameter
such as the central charge $c$ is varied, this phenomenon may occur at particular ``rational'' values
of the parameter. This point will enter the discussion repeatedly, beginning in the following 
Section \ref{sec:ent}.


\subsection{Application to entanglement subspaces of ground state}
\label{sec:ent}

Now we turn to the applications of the preceding ideas to QH trial wavefunctions. In this section
we translate the preceding results, and use some additional ideas to derive statements about the
entanglement subspaces of ground state (or other) wavefunctions obtained from a MR construction.

The starting point is to write the trial ground state wavefunction as in eq.\ (\ref{ground}), following
MR. In that work the operator ${\cal O}_{-N}$ was a uniform charge distribution, and it produces
the Gaussian factors $e^{-|z|^2/4}$ in the wavefunctions in the plane (in the limit that the
background charge distribution extends to infinity---note that this violates neutrality). These
factors are part of the lowest Landau level wavefunctions (in the symmetric gauge), and are important
when calculating quantum-mechanical inner products. In this paper we usually do not consider these
inner products, and so the Gaussian may be dropped. It is then convenient to use instead the point
charge operator
\be
{\cal O}_{-N}=e^{-iN\varphi(Z)/\sqrt{\nu}}
\ee
and place it at $Z\to\infty$. This means that (the polynomial part of) the trial ground state
wavefunction is given by the ``conformal block''
\be
\Psi(z_1,\ldots,z_N)=\langle 0_N|a(z_1)\cdots a(z_N)|0\rangle.
\label{ground2}
\ee
More generally, we may then consider trial functions that are obtained as the ``amplitudes'' we have
been discussing, that is functions in $I_N$ (or $J_N$) for the CFT under consideration:
\be
\Psi_v(z_1,\ldots,z_N)=\langle v,a(z_1\cdots a(z_N)|0\rangle\rangle.
\ee
The spaces of amplitudes $I_N$, obtained as $v$ ranges over $V^\ast$ (or $J_N$ similarly, using out states)
form spaces of nice trial wavefunctions.

The spaces $I_N$ are dual to $W_N$. Using the map $\cal Q$, we can map $W=V^0$ to $V^{-N'}$, and
similarly for $W_N$. Then Theorem 1 above implies that as $N'\to\infty$, the sum of spaces
$\bigoplus_N I_N$ can be identified as dual to $V$ itself. That is, the ground state for some $N=N'$
is viewed as the vacuum (approximate vacuum, for finite $N$), and other states, that differ from
it in charge and degree by amounts that stay constant as the limit is taken, are viewed as excited; then
finally take $N'\to\infty$. In other words, if we suppose that the
space of amplitudes can be identified as a space of zero-energy wavefunctions, and the latter can
in turn be viewed as a space of edge states, we obtain the fact familiar in examples that
the edge excitations (including those that change the particle number) of the ground state of a disk
span the (dual of the) vacuum representation of the chiral algebra (VOA) used in the construction,
as $N\to\infty$. (Such a relation of edge CFT with that used in the MR construction was proposed
in MR \cite{mr}.) In particular, the characters of the spaces agree, provided we subtract the charge
(degree) of the ground state wavefunction from the total charge (degree) before taking the
limit---this corresponds to the use of the operator $\cal Q$.
This relation is one of those sometimes referred to as a ``bulk-edge correspondence''.

Now let us consider the spaces $J_N$ of amplitudes defined using the inner product on the CFT
(for the moment, it may be the same space as $I_N$), and in particular view the ground state in
this way, as in eq.\ (\ref{ground2}). We let $N=N_A+N_B$, and consider $N_B$ of the $a(z)$s
(say, those with $i=N_A+1$, \ldots, $N_A+N_B$) as acting to the left. This can be done by taking
the absolute values of their coordinates $z_i$ larger than all those of the $z_i$s with $i=1$,
\ldots, $N_A$. Also, we may expand the former $a(z)$ in modes. Thus we arrive at out states of the form
\be
\langle0_N|\prod_{k=1}^{N_B} a_{n_k}.
\ee
Under the map (``adjoint'') from $V$ to out states, these correspond to
\be
\prod_{k=1}^{N_B} a^\ast_{-n_k}|0_N\rangle.
\ee
Comparing again with the first main result above, and replacing $a$ in the argument by $a^\ast$
(and reversing the sign of the charges), we learn that as $N_B\to\infty$, these states eventually
span $V$. This means that applying this expansion to the ground state, the ground state decomposes
into combinations of amplitudes in $J_{N_A}$, and as $N_B\to\infty$, {\em all of $J_{N_A}$ is obtained}.

This decomposition in fact involves the entanglement subspaces. Dividing the particles into those
with coordinates inside some disk of radius equal to one, centered at the origin, and those outside,
is the definition of a real-space cut. We may sharpen the preceding argument by imagining that
we insert a complete set of states in $V$,
\be
I=\sum_{v\in V} |v\rangle\langle v|
\ee
(for an orthonormal basis set of $v$) at radius one. Then we arrive by the preceding argument at
the same conclusion, now for the real-space partition entanglement subspaces in the limit as the
size of part $B$ tends to infinity.

The dimensions of the subspaces in real-space entanglement are the same as
those in particle partition. This was proved in Refs.\ \cite{drr1,sterd} by a linear-algebra argument.
In the present context, it is nicer to explain it in terms of functions \cite{rss}. The real-space
cut leads to functions in $N_A$ ($N_B$) variables restricted to lie in the disk, part $A$ (or
its complement for $B$). A function defined in the plane has a unique restriction to the disk,
and a polynomial defined on the disk has a unique analytic continuation to the plane; these maps are
inverse of one another. Now suppose
that the Schmidt decomposition is known for, say, real-space partition. The Schmidt eigenvectors can
be viewed as functions on the plane (as for particle partition), however the inner products differ, because
the integration domains are different. The decomposition for particle partition (in the plane)
can be obtained from the real-space one by re-orthonormalizing the functions in the entanglement
subspace. Hence, even though the eigenfunctions (and eigenvalues) for the two partitions differ in general,
the entanglement subspaces that they span correspond under the restriction/continuation maps between
spaces of polynomials in the plane and the disk. This is a stronger statement than simply saying that the
dimensions are equal.

Hence we have shown that as $N_B\to\infty$, the
particle (real-space) entanglement subspaces of the ground state in a MR construction coincide with
(are isomorphic to) the spaces $J_{N_A}$. This is the promised result concerning version (i)
in the Introduction. Though for $N_A\to\infty$ this result follows from the
recent work \cite{drr2}, the proof here makes no use of the quantum-mechanical inner products
on the polynomials (which required use of a generalized screening hypothesis), and consequently
applies to a larger class of constructions; moreover, the result here holds in finite sizes $N_A$.
(It may be possible to extract the $N_A\to\infty$ result for certain CFTs also from Ref.\ 
\cite{estienne2}, which does not explicitly discuss entanglement, however, the present approach is 
both more direct and, again, more general.) In the {\em orbital} partition at fixed $N_A$, the 
entanglement subspaces are generally smaller than those in the other two forms, though it is possible 
that the dimensions become the same as the limit $N_A\to\infty$ is taken.

In the preceding two results, the use of distinct spaces $I_N$ and $J_N$ in each one was necessary
in general, as we will see momentarily. But when the inner product on $V$ is non-degenerate,
the two spaces are identical. We then obtain a correspondence of the entanglement subspaces
(as $N_B\to\infty$) with the spaces of amplitudes (or zero-energy states, or edge states)
for $N=N_A$ particles. This is a finite-size form of bulk-edge correspondence.

Now we consider what happens if the inner product on $V$ is degenerate, and there are singular vectors
in $W$. The space of amplitudes $I_N$, for the given CFT in the MR construction, are to be viewed as 
spaces of ``nice'' functions by construction. For the entanglement subspaces of the ground state, it is
the inner product in the CFT that is important, and hence the spaces $J_N$. This can be seen in
several ways. For example, the use of a complete set of states as above presupposes a non-degenerate
inner product. Or we may note that the same ground state function is obtained if in the MR construction
one takes the quotient of the VOA by all singular vectors; the singular vectors do not contribute
to correlation functions in the vacuum \cite{bpz,dfms}. A particular example is the
${\cal N}=1$ superconformal model
whose wavefunctions were studied in Section \ref{scft}. The ground state wavefunction can be determined by
the superconformal algebra alone \cite{simon09}, regardless of whether the singular vectors are modded out.

This leads to an important consequence for such cases: if there are singular vectors in $W$,
the entanglement subspaces $J_N$ will generally be {\em smaller than} the corresponding space
of amplitudes $I_N$ (or zero energy states, when a special Hamiltonian exists), even as
$N\to\infty$. This is true in particular for the superconformal examples at rational central charge.
Some of the pseudoenergies (defined as minus the logarithm of the Schmidt singular values) in the
Schmidt decomposition of the ground state must
go to infinity as $c$ approaches such a value. In this case the correspondence with the spaces of
amplitudes (or ``edge states'') is lost. However, the correspondence can be recovered by using
instead the quotient CFT when finding the amplitudes or zero-energy states. This will however
involve a change in the special Hamiltonian, as we will now explain.

\subsection{Zero-energy states and special Hamiltonians}
\label{sec:appham}

In this Section, we begin to merge the streams of thought in this paper, applying the preceding analysis
of this section to the problem of solving or constructing special Hamiltonians, using the ${\cal N}=1$
SCFT model of Section \ref{scft} as a specific example. In that section, we saw that the short-distance
ope behavior of the SCFT implies that the entanglement subspace for three particles in the ground
state begins with one polynomial in degree six, followed by higher degrees, in the relative or
internal variables; thus the other degree six, and the lower ones are missing. This means that
imposing such vanishing conditions to be obeyed by the wavefunctions, or equivalently using a
three-body special Hamiltonian that enforces such conditions on zero-energy states of the
$N$-particle system, produces spaces of polynomials for each $N$ that must contain the spaces of
amplitudes of the SCFT, as obtained by the MR construction that we have discussed in this section.

It is not immediately clear whether the spaces of zero-energy states are exhausted by the spaces of
amplitudes, or whether they also contain other linearly-independent states. This issue arises
in most examples of special Hamiltonians, and we cannot always answer it. But in examples in which
we can obtain all the zero-energy states, we can compare the counting of
the states with that of the spaces of amplitudes. For the latter, we lack detailed results
for finite $N$ in general. However, it is a standard problem in CFT to compute the character
of the vacuum module, which as we have seen corresponds to the $N\to\infty$ limit of the
character of the spaces of amplitudes, and we can then compare these limits. With that information
at hand, we will be able to show the equality of the spaces even for finite $N$.

Returning to the system of Section \ref{scft}, we have emphasized that the full characters include
the counting of states at different charges from the vacuum. But it will be sufficient here to examine
the limits for a sector of fixed charge $N'-N$, as $N'\to\infty$, because the interesting part is the
$\psi$ CFT. The result in eq.\ (\ref{scft_n_infty}), with $m_3$ even, can be viewed as the sector of
zero charge excited, relative to the vacuum (the other charge-even sectors differ only by a power of
$q$, left over when removing the power $q^{\frac{3}{2}N'(N'-2)}$ for the ground state with $N'$
particles, and the charge-odd sectors also involve summing over $m_3$ odd instead of even). The
expression has factorized into $1/(q)_\infty$, which as we mentioned corresponds to excitations
of the charge sector, and two other factors, each of which is a sum. They can be expressed in a
different form as follows. We use combinatorial identities of Euler \cite{hardy}. By specializing
these general results, we can obtain for the $m_2$ sum
\be
\sum_{b=0}^\infty \frac{q^{2b}}{(q)_b}=\frac{(1-q)}{(q)_\infty}.
\ee
For the sum over $m_3$, we first leave it unrestricted, and obtain two identities,
\be
\sum_{f=0}^\infty \frac{(\pm1)^f q^{\frac{1}{2}f^2+f}}{(q)_f}=\frac{\prod_{r=1}^\infty(1\pm
q^{r-\frac{1}{2}})}{1\pm q^{1/2}}.
\ee
We may produce identities for the sums with $m_3$ even (odd) by taking half of the sum (difference) of
the expressions with the plus and the minus.

 If the simple factors on the right hand side are omitted
(that is, $1-q$ in the first, and $1\pm q^{1/2}$ in the others), these products have familiar
interpretations as the partition functions for chiral bosons and Majorana-Weyl fermions, respectively.
But with those factors, a different interpretation is more useful.

Combining the three factors in the full charge-zero character, it will be convenient again to
consider the unrestricted $m_3$ sum, weighted with $(\pm 1)^{m_3}$ (a ``${\bf Z}_2$ graded''
character). If we write $\cal V$ for the space of states in the limit with the charge sector
removed completely (the odd and even parts have to be combined carefully with the charge sectors,
as we have seen), then we have
\be
{\rm ch}^\pm_q\,{\cal V}\equiv{\rm ch}_{q,+1}\,{\cal V}\pm{\rm ch}_{q,-1}\,{\cal
V}=\frac{(1-q)\prod_{r=1}^\infty(1\pm q^{r-\frac{1}{2}}) }{(q)_\infty(1\pm q^{1/2})}.
\ee
This character can be identified as that of the Kac module for the vacuum module  of the
${\cal N}=1$ superconformal algebra (in the Neveu-Schwartz sector). This is obtained as follows.
The Verma module \cite{bpz,dfms}
consists of all formal descendants of the identity under the positive modes $L_{-n}$, $G_{-n+1/2}$
of the stress tensor and superconformal current, respectively, where $n>0$ is an integer. Using
standard arguments (the Poincare-Birkhoff-Witt basis), the Verma module ${\cal V}_0$ of the
identity is found to have the (graded) character
\be
{\rm ch}^\pm_q\,{\cal V}_0=\frac{\prod_{r=1}^\infty(1\pm q^{r-\frac{1}{2}}) }{(q)_\infty},
\ee
just as if all the modes $L_{-n}$, $G_{-n+1/2}$ for $n>0$ (anti-) commuted (although they don't).
Kac studied singular (``null'') vectors in such modules for the Virasoro and ${\cal N}=1$ superconformal
algebras, and found in particular that the vacuum module, for which the conformal weight is $0$, has a
singular vector at level $1$ in the Virasoro sector, and level $1/2$ in the superconformal current
sector, for any value of the central charge $c$. The quotient by these singular vectors, which
generate a submodule in the Verma module, leaves a module with exactly the character
${\rm ch}^\pm_q\,{\cal V}$, and it is irreducible for generic values of $c$. That is, $L_{-1}$ and
$G_{-1/2}$ annihilate the vacuum in the quotient module. (For the Virasoro part, the
fact that the vacuum is annihilated by $L_{-1}$ means it is translation invariant in the radial
quantization point of view.) This conclusion is precisely what we expected according to Theorem 1.

To avoid confusion, we should emphasize that the VOA generated by $a(z)$ and $a(z)^\ast$, which describes
the chiral algebra of the ``edge theory'' of our states (or would, if the bulk were gapped)
is not superconformal; it does not contain a superconformal current of weight $3/2$, as one may see
from the character above in the neutral sector (like all other half-odd-integer weights, it is removed
from the character ${\rm ch}^\pm\,{\cal V}$ by the projection to the even sector). Of course, the original
superconformal current $G$ still occurs, but only as part of a charged field $a$. For the basic
case $\nu=1/3$, this operator has weight $3$ and is bosonic. The chiral algebra does contain a copy of the
Virasoro algebra.

We can now draw a conclusion about the space $\widetilde{I}$ of zero-energy states of
the special Hamiltonian. We know
that it must contain the space of amplitudes $I_N$ of the VOA as we have discussed.
But the count of states shows that in fact, for any non-zero value of $c$, these spaces coincide
at infinite $N$, in a suitable sense, that involves a shift in both the charge and degree that
goes to infinity in the limit. This implies that the spaces $I_N$ and $\widetilde{I}_N$ are the same
in a certain limit of large $N$ and large degree $d$. But we can now prove
\newline
{\bf Theorem 2}: In the example of this section, the spaces of polynomials $I_N$ (the amplitudes)
and $\widetilde{I}_N$ (the zero-energy states) are the same for any particle number $N$.
\newline
The detailed proof, which uses only arguments about symmetric polynomials (plus some second quantization
for convenience only) is given in Appendix \ref{app:thm2}. The idea of the proof is to show that if there
are any more vanishing conditions defining $I_N$, or elements generating $\cal I$, then functions
are removed by
these conditions from $\widetilde{I}_N$ at all $N$ and at fixed difference in degree from the ground
state value. Then the coefficients in the character series become smaller in the limit, and this
occurs in all charge sectors in $V$. This contradiction with the previous results completes the proof.
Hence the ideal $\cal I$ is generated by ${\cal I}_3$. The characters of the components of the
principal subspace are ${\rm ch}_q\,W_N={\rm ch}_q\,\widetilde{I}_N$, which we have calculated. (For
$W_N$, we define the character unconventionally, as ${\rm ch}_q\,W_N={\rm tr}_{W_N}\,q^{L_0-Nh}$.)

We want to emphasize that the technique used to prove Theorem 2 is very general. It can
be applied to other examples (for example, the Laughlin states \cite{laugh}, the Read-Rezayi (RR) series
\cite{rr99}, and the Gaffnian \cite{srcb}, as well as the examples analyzed in Section \ref{sec:M3p})
in which some translation-invariant relations involving
some number of $a$'s and their derivatives are known, or equivalently given a special Hamiltonian,
if two conditions are met. These are that we know the zero-energy states in each degree $d$ and particle
number $N$, and that these numbers agree with the vacuum module $V$ of the CFT (or VOA) asymptotically
in the standard limit for dealing with the edge states. When these conditions are met, the equality
then extends to finite $N$. We point out that it has been a perennial problem when using the MR
construction to show that the zero-energy states of some special Hamiltonian are precisely the
spaces of amplitudes, and no larger (see e.g.\ Ref.\ \cite{rr99}). For the case of the RR states,
the proof was constructed earlier \cite{fs}, but to the present authors it appears technically
difficult, and it may not be possible to extend it to other cases. By contrast, the argument given
here deals with the MR and RR states using
only the counts of states that have long been known. In FJM, the same result is reached in their
examples, but the last part of the argument is not explicitly given. Our approach also differs in
that we have methods to find all the
zero-energy states explicitly, and not only to give an upper bound. These methods were completed
for the RR states by the authors of Ref.\ \cite{aks}, but they quoted the old result of Ref.\
\cite{fs} to obtain the equality of the spaces of polynomials, rather than obtaining that as a
byproduct of the construction.

Next we turn to the inverse problem, finding special Hamiltonians such that the zero-energy subspaces
are some desired spaces of functions, such as the amplitudes $I_N$ or $J_N$ of a given MR construction.
This is essentially the problem of finding ``vanishing conditions'' that characterize these spaces.
If the spaces of amplitudes $I_N$ or $J_N$ can be defined as symmetric polynomials that satisfy some
``vanishing conditions'' as $z_i$s come to the same point, we may construct a Hamiltonian that
annihilates these functions and gives positive eigenvalues to
functions not in the subspace, by forming projection operators that project onto functions (or states
in the quantum mechanical Hilbert space, of lowest LL states of $N$ particles) that are {\em not} in
the subspace. Now the amplitudes form a subspace, and functions linearly independent from them lie in
the quotient space by that subspace. There is no natural subspace of representatives for this
quotient space, in the absence of an inner product. In the quantum mechanical situation, we may use
the $L^2$ inner
product on lowest LL states of the plane, sphere, or torus, depending on our own interest. Hence the
form of the projection operator, onto some functions (in $N'=2$, $3$, \ldots variables) linearly
independent from the ``allowed'' subspace, necessarily depends on the system geometry, and even on
its size ($N_\phi$). Hamiltonians that are a sum of such terms (necessarily invariant under
permutations) with positive coefficients (the magnitude of the coefficients is not important for
our discussion) have been in use in the QH effect for some time (see e.g.\
\cite{hald83,gww,rr96,rr99}). Because the relations to be imposed on the functions are local, the
projection operators in these special Hamiltonians are short range, even if they involve interactions
among more than two particles.

In the known examples, the interactions represented by the projection operators involve a
bounded number of particles, and bounded degree for the relations (total number of derivatives
appearing in the vanishing conditions) for each number of particles. If these conditions are not met,
then the special Hamiltonian would not be local. (Higher degrees, which must necessarily occur
if more particles are involved, imply a more extended interaction). Actually, this need not be
strictly true. We only require the projection operators to have positive coefficients, while the
ground and quasihole states would be of zero energy. The higher degree, and so longer range,
terms could be given numerically smaller coefficients, and then the Hamiltonian can be
relatively short range (or effectively involve power laws in distance). But it seems preferable
in aiming for physical examples that the boundedness in particle number and degree of the terms
be respected as far as possible.

Because of the duality discussed in Sec.\ \ref{sec:dual}, we can describe a set of spaces of
polynomials (one space
for each $N$) in terms of the algebra $\cal A$ and and a translationally-invariant ideal $\cal I$,
with the action of the positive currents modes (and hence, symmetric polynomials) and the translation
generator as before, even in the absence of a MR construction from a CFT. (It is less clear if
we should also
insist on an action of the positive half of the Virasoro algebra.)  Then the
space of polynomials that satisfy the vanishing conditions obeys the duality
$\widetilde{I}\cong{\cal A}/{\cal I}$. This is the point of view we will take for a while.

In terms of the discussion of the spaces $\widetilde{I}$, and the relations
${\cal I}$ in $\cal A$, this means that the former are determined by a set of translation-invariant
vanishing conditions, that should preferably be bounded in both $N_A$ and $d$, and consequently finite.
In terms of the relations in the ideal $\cal I$, this property admits a clear mathematical definition:
there is a {\em finite} set of elements $i_l$ (indexed by $l$), each $i_l$ lying
in a subspace ${\cal I}_{d_l,N_l}$, such that $\cal I$ is generated from $\{i_l\}$ by taking
linear combinations with coefficients in $\cal A$ {\em and} using the action of positive current modes,
and the translation generator $L_{-1}$. (We should point out that $L_{-1}$ acts on $\xi_{-n}$ in $\cal A$
in the same way as on $a_{-n}$, except for the shift in index by $h$, and so raises the total
degree of excitation $\sum_i n_i$, as it is the transpose of the derivative operator in the dual space
of symmetric polynomials.)
It is far from clear if this finiteness property will generally be the case for the VOAs of
interest here. This question now becomes an important issue to be studied in future work.

To describe the structure of the special Hamiltonians in practice, some additional notation will be useful.
To present the information about what terms appear in the special Hamiltonian, we can form a character
for the part involving interactions among $N_A$ particles, in which the coefficients are the dimensions
of the space onto which the projection operators project in each degree. For example, for
the special Hamiltonian for the generic SCFT model, the series is
\be
1+q^2+q^3+q^4+q^5+q^6
\ee
for the three-body terms, and zero for $N_A\neq 3$. This series is the character of the space
${\cal I}_3$, omitting the factor $(1-q)^{-1}$ which describes the uninteresting and ubiquitous center
of mass degree of freedom that relates to translation invariance. Comparing with the series in
Eq.\ (\ref{ch_N3})
for the internal part for the space of all symmetric polynomials, we see that the first term allowed is
one of those in degree six, followed by all those of higher degree. As these series are indeed finite at
least in the examples we will discuss, it is simpler to present this information than the character
for the ``allowed'' functions in $\widetilde{I}$. (Recall that ${\rm ch}_q\,\widetilde{I}_3={\rm
ch}_q\,\Lambda_3 - {\rm ch}_q\, {\cal I}_3$.)
Thus this type of series allows a succinct description of the terms in the Hamiltonian, though
without giving the precise vector onto which projection is performed,
which is important for the complete description when there is more than one, as occurs here for degree
six for three particles. We seek to identify a set of generators (in the sense above) for $\cal I$,
and the use of a corresponding term in the Hamiltonian for each one should produce as the zero-energy
subspace $\widetilde{I}_N$ for all $N$.

Because we consider finite sets of generators, it makes sense to add generators one by one, that is to
add additional translation-invariant terms to the special Hamiltonian. In addition
we must preserve the property that $\cal I$ is a module over the (dual action of the) symmetric polynomials
or positive current modes, if we are to have any hope of finding polynomials that are zero-energy states
for all $N$, or of agreeing with some MR construction from CFT. On $\cal I$, the current modes $j_n$
act by lowering the excitation degree, as they are the transposes of the sums of powers $s_n$ that raise
the degree of the symmetric polynomials.

The special Hamiltonian for the SCFT model above, like other known simple examples, satisfies
these properties in a basically trivial way. Further cases will require less trivial checks that the
ideal is a module over the positive current modes. To find examples, we return to the observation that
the space of amplitudes $J_N$, constructed from inner products, which correspond to the entanglement
subspaces of the (very large) ground state, may be smaller than the space of amplitudes $I_N$
constructed from the dual space, when the latter contains singular vectors. The ``missing'' vectors
in $J_N$ give us an opportunity to add elements to $\cal I$, so that the missing vectors are those
forbidden by the additional vanishing conditions.

For the ${\cal N}=1$ SCFT, it is known \cite{kac} for
\be
c=
\frac{3}{2}\left(1-\frac{2(p-p')^2}{pp'}\right)
\label{c_rational}
\ee
where $p$, $p'$ are integers, $p-p'$ is even, and $p$ and $(p-p')/2$ are coprime, that there
are additional singular vectors in the modules,
in particular in the vacuum module. This implies that they will occur at some corresponding degree
in the spaces $I_N$ also, at least for sufficiently large $N$. Setting all singular vectors to zero
results in the superconformal minimal models, denoted $SM(p,p')$. One case is the tricritical
Ising model example,
which however still seems to be difficult. Here we will pursue a simpler example, studied in FJM (though
not from the SCFT point of view). Like the tricritical Ising model, the $M(3,8)$ Virasoro minimal
model (which has central charge $c=-21/4$) contains the ${\cal N}=1$ superconformal current at a corner
in the Kac table, $(2,1)$ in the
$M(3,8)$ case, and hence $M(3,8)=SM(2,8)$. FJM showed that in this case, the ideal for three
particles has the character
\be
(q)_1{\rm ch}_q\,{\cal I}_3=1+q^2+q^3+q^4+q^5+q^6 +q^8,
\ee
and that ${\cal I}_3$ generates $\cal I$. The corresponding vanishing conditions thus determine spaces
$\widetilde{I}_N=I_N$ of polynomials for each $N$ which are those of the VOA of the SCFT model,
with the singular vectors set to zero. Compared with the special Hamiltonian of the generic
superconformal model, an additional translation-invariant term has been added in degree 8,
for three particles, and nothing else. In the next section, we analyze this model, and some others,
in detail. These models provide non-trivial examples of the structure we have just discussed.

\section{Analysis of special Hamiltonians for $M(3,p)$ series}
\label{sec:M3p}

In FJM \cite{fjm}, the example described at the end of the previous section was analyzed in detail.
It is just one
in a series in which the CFT used is the $M(3,p)$ minimal model, $p=4$, $5$, $7$, \ldots ($p$ is not
divisible by $3$), with the central charge of the $M(3,p)$ minimal model
\be
c=1-\frac{2(p-3)^2}{p}.
\ee
The field $\psi$ is the $(2,1)$ field in the Kac table, with conformal weight
$h_\psi=(p-2)/4$. We concentrate on the basic cases of bosons, for which the filling factor is
again $\nu=1/(2h_\psi)$. These examples of the MR construction have also been described as $(k,r)=(2,p-2)$
models; here as before $k=2$ means they are paired states, while $r=p-2$ is the degree of the lowest
non-vanishing allowed behavior for $k+1=2$ variables.

The analysis of FJM begins with the three-particle ground state wavefunction (symmetric polynomial)
they call $\varphi_3$. They show that the space $I_3$ is generated over the symmetric polynomials
by $\varphi_3$ and $\ell_1(\varphi_3)$, and compute the character, as presented above for $M(3,8)$.
Taking the three-particle ideal
${\cal I}_3$ as determining vanishing conditions, they then analyze the zero-energy subspaces for
all $N$. Because a bound on the characters of these agrees asymptotically with a so-called fermionic
character formula derived from that for the $M(3,p)$ minimal model, they conclude that ${\cal I}_3$
generates $\cal I$, and $\widetilde{I}_N=I_N$ for all $N$. They also calculate the characters for
functions on the sphere, corresponding to counting quasihole states.

We will not repeat more details of FJM's analysis here. That is because the remaining methods do not
obviously generalize. We do two things: we construct explicit sets of wavefunctions for all the
zero-energy states, along similar lines as above for the generic superconformal model; and we describe
the three-body interaction Hamiltonian for which these are the zero-energy states, for both the plane
and the sphere. For completeness, we do the same for the other members of the $M(3,p)$ sequence.

\subsection{$M(3,8)$ wavefunctions in the plane}

To find the wavefunctions, we make use of FJM's results. The vanishing conditions on the $N$-variable
functions are incorporated by requiring that any function occurring in the particle-partition
decomposition for any $N_A=3$ variables must lie in the span of $\varphi_3$
 and $\ell_1(\varphi_3)$ times symmetric polynomials, that is in $\widetilde{I}_3=I_3$. They use
 a filtration obtained as in Sec.\ \ref{scft}
by setting variables equal in pairs. A key point is that in the present case, because of the additional
vanishing condition, the kernel of this map applied to three particles lies not only in $D_N
\widetilde{I}_N^{\rm MR}$, it lies in its subspace $D_N \widetilde{I}_N^{\rm Gf}$, where
$\widetilde{I}_N^{\rm Gf}$ is the space of symmetric
polynomials that vanish as degree three or higher when any three variables coincide (the ground state
or lowest degree polynomial in this space has been called the Gaffnian function \cite{srcb}). These
are related to the $M(3,5)$ minimal model, and are again explicitly known \cite{srcb}.

Combining this observation with the construction of functions as before, we find a complete
and linearly-independent set of polynomials in $\widetilde{I}_N^{[8]}$, essentially by replacing
elements of the ideal $\widetilde{I}^{\rm MR}_{N-2m_1}$, that was used before, with elements of
$\widetilde{I}^{M(3,5)}_{N-2m_1}$. Explicitly, with notation
as before, we replace (4.9) with
\begin{multline}
\label{eq_sm28_plane_basis}
\calS_z \Biggl\{ \prod_{(1,i)>(1,i')} \chi( z_{i,1}^{(1)}, z_{i,2}^{(1)} ; z_{i',1}^{(1)},
z_{i',2}^{(1)})\cdot
\Biggr. \\
\times\prod_{\alpha'=2,3} \prod_{(1,i)>(\alpha',i')} \prod_{j=1}^{r_{\alpha'}} \chi_3 ( z_{i,j}^{(1)},
z_{i,j+1}^{(1)} ;
z_{i',j}^{(\alpha')})\cdot \\
\times \prod_{(2,i)>(\alpha',i')} \prod_{j=1}^{r_{\alpha'}} \chi^\text{Gf}_3 ( z_{i,j}^{(2)},
z_{i,j+1}^{(2)} ; z_{i',j}^{(\alpha')})\cdot \\
\times \prod_{(3,i)>(3,i')} ( z_{i,1}^{(3)} - z_{i',1}^{(3)} )^2\cdot \\
\Biggl. \times D_{N-2m_1} \! \left( \left\{ z_{i,j}^{(\alpha = 2,3)} \right\} \right)\cdot
\prod_{\alpha} \prod_{l=1}^{m_\alpha} e_l^{n_l^{(\alpha)}} \! \! \left( \left\{ Z^{(\alpha)}_i \right\}
\right) \Biggr\}.
\end{multline}
Here the $(\alpha,i)$th cluster center of mass is $Z^{(\alpha)}_i = \frac{1}{r_\alpha}
\sum_{j=1}^{r_\alpha}
z_{i,j}^{(\alpha)}$, and the only difference from the generic SCFT case is in the third line, where we
define
\begin{align}
\label{eq_gf_chi3}
\chi_3^\text{Gf}(z_1,z_2; z_3)& = (z_1-z_3)^2(z_2-z_3),\\
\label{eq_gf_chi4}
\chi^\text{Gf}(z_1,z_2; z_3,z_4)& = \chi_3^\text{Gf}(z_1,z_2; z_3)\chi_3^\text{Gf}(z_2,z_1; z_4).
\end{align}
Note that this only affects couplings between pairs of particles of type $\alpha =2$, and between a pair of
type two and a particle of type three.

The degree of each function, omitting the elementary symmetric polynomial factors, is now
\be
\frac{3}{2}N(N-2) + m_2^2 + m_2 m_3 + \frac{1}{2}m_3^2+m_2+m_3.
\ee
Hence the character is identical to that given in FJM:
\begin{multline}
\label{eq_m38_finite_char}
q^{-\frac{3}{2}N(N-2)}\text{ch}_q \widetilde{I}^{[8]}_N =  \\
\sum_{\substack{m_2,m_3 \geq 0: 2m_2 +m_3 \leq N, \\ (-1)^{m_3} = (-1)^N}} \frac{q^{m_2^2 + m_2 m_3 +
\frac{1}{2}m_3^2+m_2+m_3}}{
(q)_{\frac{N-2m_2-m_3}{2}} (q)_{m_2} (q)_{m_3}}.
\end{multline}
We emphasize that these describe the dimensions of the entanglement subspaces for the ${\cal N}=1$ SCFT
ground state for $c=-21/4$.

For $N\to \infty$ we can add the odd and even sectors to recover the $SM(2,8)$ vacuum character in
fermionic form \cite{temp_sm_chars1}
\begin{align}
(q)_\infty {\rm ch}_q {V}_{(1,1)}^{SM(2,8)} &= (q)_\infty \left( {\rm ch}_q{ V}_{(1,1)}^{M(3,8)} + q^{3/2}
{\rm ch}_q{ V}_{(2,1)}^{M(3,8)}
\right) \nonumber \\
\label{eq_m38_char}
{}&= \sum_{m_2,m_3 \geq 0} \frac{q^{m_2^2 + m_2 m_3 + \frac{1}{2}m_3^2+m_2+m_3}}{(q)_{m_2} (q)_{m_3}},
\end{align}
the first few terms of which are
\begin{align}
(q)_\infty {\rm ch}_q{ V}_{(1,1)}^{M(3,8)} &= 1+q^2+q^3+2q^4+2q^5+4q^6+\ldots,\\
(q)_\infty {\rm ch}_q{ V}_{(2,1)}^{M(3,8)} &=
q^{3/2}(1+q+q^2+2q^3+3q^4+4q^5+\ldots).
\end{align}
These should be compared with those for the vacuum Kac module of the generic case.

\subsection{$M(3,8)$ wavefunctions on the sphere}

As for the generic superconformal model, the wavefunctions on the sphere can be found with only
slightly more effort. The degree in each coordinate variable is found to be $\np
= (6N+n)/2-6$, as this is determined only by the ``leading order''/``ground state'' clustering behavior
parameters $(k =2,r=6)$. Repeating the power counting in the residues of the wavefunctions
\eqref{eq_sm28_plane_basis}, we find that the required number of quasiholes of each type are
\begin{align}
\label{eq_m38n1}
n_1 &= n, \\
\label{eq_m38n2}
n_2 &= n-2m_2-m_3-2, \\
\label{eq_m38n3}
n_3 &= \frac{n}{2}-m_2-m_3-2.
\end{align}
The factors involving particles of type one are unchanged; we only need to homogenize the coupling
\eqref{eq_gf_chi3} between a half-broken pair of type $\alpha=2$ and a particle of type three:
\be
\chi_3^\text{Gf}(z_1,z_2; z_3; w_1) = (z_1-z_3)^2(z_2-z_3)(z_1-w_1)(z_2-w_1)^2.
\ee
Again, because $n_2 = 2n_3+m_3+2$, there are always enough quasiholes of type two to do this. The basis
functions on the sphere are therefore
\begin{multline}
\calS_z \Biggl\{ \prod_{(1,i)>(\alpha,i')} \chi( z_{i,1}^{(1)}, z_{i,2}^{(1)} ; z_{i',1}^{(\alpha)},
z_{i',2}^{(\alpha)}) \Biggr. \\
\times \prod_{(1,i)>(3,i')} \chi_3 ( z_{i,1}^{(1)}, z_{i,2}^{(1)} ; z_{i',1}^{(3)}; w^{(1)}_{2i'-1},
w^{(1)}_{2i'})  \\
\times \prod_{(2,i)>(2,i')} \chi^\text{Gf} (z_{i,1}^{(2)}, z_{i,2}^{(2)}; z_{i',1}^{(2)}, z_{i',2}^{(2)})
\\
\times \prod_{(2,i)>(3,i')}
\chi_3^\text{Gf} (z_{i,1}^{(2)}, z_{i,2}^{(2)}; z_{i',1}^{(3)}; w_{i'}^{(2)}) \\
\times \prod_{(3,i)>(3,i')} ( z_{i,1}^{(3)} - z_{i',1}^{(3)} )^2  \cdot D_{N-2m_1} \! \left( \left\{
z_{i,j}^{(\alpha = 2,3)} \right\} \right)\\
\Biggl. \times
\prod_{\alpha} \prod_{i,j} \sideset{}{^{(\alpha,j)}}\prod_{l} (z_{i,j}^{(\alpha)} - w_l^{(\alpha)})
\Biggr\}.
\end{multline}
The ranges of the quasihole factors in the last line are
\begin{flalign}
l &= 2m_3+1, &\ldots,  &2m_3+n_3+2, &(\alpha=1, j=1), \nonumber \\ l &= 2m_3+n_3+3, &\ldots, &n_1,
&(\alpha=1, j=2), \nonumber \\ l &= m_3+1, &\ldots, &(n_2+m_3)/2, &(\alpha=2, j=1), \nonumber \\ l &=
(n_2+m_3+2)/2, &\ldots, &n_2,  &(\alpha=2, j=2), \nonumber \\ l &= 1, &\ldots,  &n_3, &(\alpha=3, j=1).
\end{flalign}

It may be verified that the character on the sphere is still of the form
\be
\text{tr } q^{-L_z} = \prod_\alpha \binom{ m_\alpha + n_\alpha }{ m_{\alpha}}_q,
\ee
with the $n_\alpha$ now given by \eqref{eq_m38n1}--\eqref{eq_m38n3}; this is equivalent to theorem 5.14 in
FJM (where the variables $\np+1$, $N$ are denoted $N$, $n$). Taking the
planar limit $n \to \infty$ with $N$ fixed, we recover \eqref{eq_m38_finite_char}.
Furthermore, the quasihole multiplicity for fixed positions is now
\be
\sum_{\substack{m_2,m_3 \geq 0:\\N-2m_2-m_3 \geq0, \\(-1)^{m_3} = (-1)^{N}.}} \binom{n-m_2-m_3-2}{ m_2}
\binom{n/2 -m_2-2}{ m_3},
\ee
the domain of which is clearly bounded and independent of $N$ for $N>n$ (hence the sum itself also
has these properties). Even though this multiplicity is now bounded, unlike in the generic
superconformal case, we still expect this model to be gapless in the thermodynamic limit because it
is a non-unitary CFT.

\subsection{$M(3,p)$ series: general $p$}

As found by FJM, the structure discussed above for the $M(3,8)$ case holds for all $M(3,p)$ models
with $p>5$. That is, the kernel of each map $C_1$ for $\widetilde{I}_N^{[p]}$ is
$D_N\widetilde{I}_N^{[p-3]}$ (and note that $\widetilde{I}_N^{[4]}=\widetilde{I}_N^{\rm MR}$).
Then the spaces of zero-energy states for the $M(3,p)$ models can be found recursively for all $p$
not divisible by $3$. Defining
\be
s = \left \lfloor \frac{p}{3} \right\rfloor,
\ee
there will be $\alpha = 1, \ldots, s$ types of particles with $r_\alpha =2$, representing pairs in various
stages of ``brokenness,'' and one type of unpaired particles with $\alpha = s +1$, $r_\alpha =1$.
Schematically, the plane wavefunctions take the form
\begin{multline}
\label{eq_m3p_plane_basis}
\calS_z \Biggl\{ \prod_{\alpha =1}^{s} \Biggl[ \prod_{(\alpha,i)>(\alpha,i')} \chi^{(\alpha)}(
z_{i,1}^{(\alpha)}, z_{i,2}^{(\alpha)} ; z_{i',1}^{(\alpha)}, z_{i',2}^{(\alpha)}) \Biggr. \Biggr. \\
\times \prod_{(\alpha,i)>(\alpha',i')} \prod_{j=1}^{r_{\alpha'}} \chi_3^{(\alpha)} ( z_{i,j}^{(\alpha)},
z_{i,j+1}^{(\alpha)} ; z_{i',j}^{(\alpha')}) \\
\times \Biggl. D_{N-2 \sum_{\beta=1}^\alpha m_\beta} \! \left( \left\{ z_{i,j}^{(\beta = \alpha+1, \ldots,
s+1)} \right\} \right) \Biggr] \\
\times \prod_{\alpha} \prod_{l=1}^{m_\alpha} e_l^{n_l^{(\alpha)}} \! \! \left( \left\{ Z^{(\alpha)}_i
\right\} \right) \Biggr\}.
\end{multline}
All pairs of rows of type $1 \leq \alpha \leq s$ are coupled by a function
\begin{multline}
\label{eq_m3p_chi4}
\chi^{(\alpha)} (z_1,z_2; z_3,z_4) = \\
\sum_{j=1}^{\left\lceil \frac{p-3\alpha}{2} \right\rceil} c^{(\alpha)}_j \bigl( (z_1-z_3)(z_2-z_4)
\bigr)^{p-3\alpha+1-j} \bigl((z_1-z_4)(z_2-z_3) \bigr)^j,
\end{multline}
where the $c^{(\alpha)}_j$ are constants which are partially fixed by the requirement that the
symmetrization of this function reproduce the four-point conformal block:
\begin{multline}
\label{eq_m3p_chi4sym}
\calS_z \chi^{(\alpha)} (z_1,z_2; z_3,z_4) \\
=\bigl \langle a(z_1) \cdots a(z_4) \bigr\rangle_{M(3,p-3(\alpha-1))\otimes U(1)}.
\end{multline}
Because these are Virasoro minimal models, the four-point block may be evaluated in terms of the ordinary
hypergeometric function as \cite{fjm}
\begin{multline}
\calS_z \chi^{(\alpha)}(z_1,z_2; z_3,z_4) = \bigl( (z_1-z_3)(z_2-z_4) \bigr)^{p-3\alpha+1} \\
\times {}_2F_1 \! \left( \alpha -\frac{p}{3}, 3\alpha-p-1; 2\alpha-\frac{2p}{3}; x  \right),
\end{multline}
where the cross-ratio $x = (z_1-z_4)(z_2-z_3)/(z_1-z_3)(z_2-z_4).$ The leading-order behavior in $z_1, z_2,
z_3$ as $z_4 \to \infty$ then defines $\chi^{(\alpha)}_3 (z_1,z_2; z_3)$.

We use the term ``schematic'' in referring to the wavefunctions \eqref{eq_m3p_plane_basis} because
$\chi^{(\alpha)}$ is not uniquely determined by Eq.\ \eqref{eq_m3p_chi4sym}: there exist
non-symmetric functions of the form
\eqref{eq_m3p_chi4} which vanish under symmetrization but whose contributions do \emph{not} vanish when
used to construct a wavefunction of the form \eqref{eq_m3p_plane_basis} with $N >4$ particles. The
ambiguity can of course be resolved by an examination of the singularity structure of the four-point block
$\langle \psi(z_1)\cdots \psi(z_4)\rangle_{M(3,p-3(\alpha-1))}$, as was done for $G$ in
Ref.\ \cite{simon09}, but this must be done on a case-by-case basis as the conformal weight
of $\psi$ (and hence the number of singular terms in the $\psi \cdot \psi$ OPE) increases with
$p$; therefore this is the most we can say about the form of the wavefunctions without
restricting ourselves to specific cases. The $p=7$ case is the simplest; then we can use
\be
\chi_3(z_1,z_2;z_3)=(z_1-z_3)^3(z_2-z_3)^2
\ee
and
\be
\chi(z_1,z_2,z_3,z_4)=(z_1-z_3)^3(z_2-z_3)^2(z_2-z_4)^3(z_1-z_4)^2,
\ee
which resemble what was used in the Gaffnian functions. When symmetrized, $\chi_3$ gives
the unique translationally-invariant function in three variables of degree 5, and so is correct.

For general $p$, characters on the plane and sphere may still be computed without explicit basis functions;
this was done in section 5 of FJM. We restate their results in a form more conventional in the FQHE
literature \cite{temp_qholes,read06}: let
\be
\qdegen{n}{F}_p
\ee
denote the quasihole degeneracy at fixed positions in the $M(3,p)\otimes U(1)$ theory on the sphere
with $n$
quasiholes and $F$ not-fully-paired particles (in the notation of \eqref{eq_m3p_plane_basis},
$F=\sum_{\alpha=2}^{s+1} r_\alpha m_\alpha$). The relationship between $N, n$ and $\np$ is
\be
N_\phi^{[p]} = \frac{p-2}{2} N + \frac{n}{2} - (p-2).
\ee
FJM's equations (5.5), (5.6) then give a recursive expression for the quasihole degeneracy,
analogous to the result obtained in Ref.\ \cite{read06} for $\Zs_k$ parafermions:
\be
\qdegen{n}{F}_p = \sum_{ F'\equiv F \text{ mod } 2} \binom{\frac{F-F'}{2}+n'}{n'} \qdegen{n'}{F'}_{p-3}
\ee
with $n'=n-F-2$ and initial condition
\be
\qdegen{n}{F}_p = \binom{\frac{n}{2}+(p-4)\left(\frac{F}{2}-1\right)}{F}, \text{ for $p=4,5$}.
\ee
FJM solve this recursion in their theorem 5.14; in particular the numbers of quasiholes at each stage are
\be
n_{\alpha}^{[p]} = n -2(\alpha-1)-\sum_{\beta=1}^{s+1} \bigl(\min(\alpha,\beta) -1 \bigr) r_\beta m_\beta
\ee
for $1 \leq \alpha \leq s$ and
\begin{multline}
n_{s+1}^{[p]} = \frac{n}{2}-s -\frac{1}{2}\sum_{\beta=1}^{s+1} (\beta-1)r_\beta m_\beta \\
+\frac{1}{2}\bigl((p \text{ mod } 3 ) -2 \bigr)(m_{s+1}-2).
\end{multline}

\subsection{Explicit special Hamiltonians on the sphere and the plane}

So far we specified the allowed behavior for any three particles as that belonging to the spaces
$I_3^{[p]}$ for three particles. As these functions are explicitly known, we have only to
construct translation-invariant projection operators onto the orthogonal complement of these spaces,
and the sum of these any positive coefficients gives a special Hamiltonian that produces the allowed
functions as zero-energy states. Unlike the allowed functions, the forbidden behavior, which must be
orthogonal to the allowed, depends on the geometry of the system through the inner product in use.
We give results for the sphere, for which the measure in stereographic coordinates (by mapping to the
plane) is
\be
\mu(z) \, d^2z = \frac{dz \, d\overline{z}}{(1+|z|^2/4)^{2(1+\np/2)}}.
\ee
The results for the plane may be obtained by keeping only the leading order terms as $\np
\to \infty$.

FJM's results determine the character of the ideal ${\cal I}_3^{[p]}$ for the $M(3,p)$ case as
\be
(q)_1{\rm ch}_q\,{\cal I}_3^{[p]}=\frac{(1-q^{p-2})(1-q^{p-1})}{(1-q^2)(1-q^3)}.
\ee
The series on the right-hand side terminate with a finite number of terms. The coefficients of these
terms tell us how many orthogonal states must be forbidden in the polynomial behavior for each $p$.

As a basis, we use the translationally-invariant symmetric polynomials $\{ \widetilde{e}_n \}$, $2 \geq n
\geq N$, defined in Ref.\ \cite{src} in terms of the elementary symmetric polynomials $e_n$ as
\be
\widetilde{e}_n \left( \left\{ z_i \right\} \right) = e_n \biggl( \biggl\{ z_i - \frac{1}{N} \sum_{j=1}^N
z_j \biggr\} \biggr).
\ee
We specify the orthogonal allowed and the forbidden behavior for each degree $d$ in which the projection
operator has non-zero image and kernel among the functions of that degree, since otherwise an
arbitrary basis for that degree can be used (or omitted entirely). Recall that the lowest allowed degree
is $r=p-2$. The functions have not been normalized. The first examples are as follows:

$p=7$ ($r=5$): At $d=6$, the allowed function is $9\tee^2 - 2 \te^3$; the forbidden (to be projected onto
in the special Hamiltonian) is
\begin{multline}
\label{eq_m37_ham}
27\left( 9 \np^3 -48 \np^2 +73 \np -40 \right) \tee^2 \\ +\left(63 \np^3 -471 \np^2 + 1156 \np -880 \right)
\te^3.
\end{multline}

$p=8$ ($r=6$): At $d=6$, allow $\varphi_3^{[8]} \propto 9 \tee^2- \te^3$; forbid
\begin{multline}
27\left( 9 \np^3 -60 \np^2 +127 \np -100 \right) \tee^2 \\
 + 2\left(45 \np^3 -408 \np^2 + 1199 \np -1100 \right)\te^3.
\end{multline}
The behavior at $d=7$ is allowed, and at $d=8$ allow $\te \varphi_3^{[8]} \propto  9
\tee^2 \te- \te^4$; forbid
\begin{multline}
54\left( 9 \np^3 -69 \np^2 +160 \np -140 \right) \tee^2 \te \\ + \left(99 \np^3 -948 \np^2 + 2999 \np -3010
\right)\te^4.
\end{multline}

$p=10$ ($r=8$): At $d=8$, allow $\varphi_3^{[10]} \propto
-18 \tee^2 \te + \te^4$; forbid
\begin{multline}
27\left( 9 \np^3 -93 \np^2 +316  \np -392 \right) \tee^2 \te \\ + \left(63 \np^3 -786\np^2 +3217 \np -4214
\right)\te^4.
\end{multline}
At $d=9$, allow $(\ell_1 - 16/3 s_1)\varphi_3^{[10]} \propto 3 \tee^3 -2 \tee \te^3$; forbid
\begin{multline}
27\left( 9 \np^3 -84 \np^2 +235  \np -216 \right) \tee^3 \\ + \left(63 \np^3 -804\np^2 +3397 \np -4392
\right)\tee \te^3.
\end{multline}
At $d=10$, allow $\te \varphi_3^{[10]} \propto -18 \tee^2 \te^2 + \te^5$; forbid
\begin{multline}
135\left( 9 \np^3 -102 \np^2 +373  \np -504 \right) \tee^2 \te^2 \\ + 2\left(117 \np^3 -1542\np^2 +6745 \np
-9576 \right) \te^5.
\end{multline}
Both behaviors at $d=11$ are allowed, and at $d=12$ we have a three-dimensional space;
we must allow the two-dimensional space spanned by the (nonorthogonal) behaviors $ \tee(\ell_1 - 16/3
s_1)\varphi_3^{[10]} \propto 3 \tee^4 -2 \tee^2 \te^3$ and $ \te^2 \varphi_3^{[10]} \propto -18 \tee^2
 \te^3
+ \te^6$, and forbid the orthogonal complement
\begin{multline}
3645\left(81   \np^6 -1755  \np^5 +14607  \np^4 \right.\\
\left. -60117   \np^3 +134284   \np^2 -169756   \np + 110880 \right) \tee^4 \\
+270 \left(1053 \np^6 -30834 \np^5 +369873 \np^4 -2330376 \np^3 \right.\\
\left. +8190988  \np^2 -15458176 \np + 12418560 \right) \tee^2 \te^3 \\
+2   \left(20007\np^6 -641979\np^5 +8479917\np^4 -58925481\np^3 \right.\\
\left. +226850956\np^2 -458053660\np + 378655200\right) \te^6.
\end{multline}

None of the coefficients above factors as a polynomial in $\np$, and using different bases for the
polynomials [such as Jack polynomials at the appropriate parameter value $\alpha = -(k+1)/(r-1)$] does not
yield noticeable patterns in the expressions involved.

\section{Conclusion}

To conclude, we have explored the relations among several spaces of wavefunctions that can be constructed
for a system of trial wavefunctions, especially when those can be constructed from a CFT. These were: 1)
the entanglement subspaces of the ground state, for $N_A$ particles out of $N$; 2) spaces of amplitudes
(conformal blocks) constructed from a CFT; and 3) spaces of zero-energy states of a special or projection
operator Hamiltonian. In favorable cases, we can show that all of these consist of exactly the
same wavefunctions. Even when they do not, it may be possible to add terms to the Hamiltonian and
restore agreement, in terms of a ``quotient'' CFT. This is a finite-size version of a bulk-edge
correspondence, in current parlance. The techniques developed in this paper can be applied to
many other examples, and while we concentrated on wavefunctions for spinless (or spin-polarized)
particles, they also can be applied to wavefunctions for particles with spin.

We have not addressed the issue of the energy gap above the zero-energy states for the special
Hamiltonian. It is expected that when the CFT is non-unitary or non-rational, the model will be gapless
in the thermodynamic limit \cite{read09}. We want to remark here that our analysis makes it clear that
there are many terms (involving many-particle interactions) that can be added to such a Hamiltonian,
such that the zero-energy states are unchanged. The non-zero energy states would be affected, however.
But it should be the case that no such terms can produce a gap in the bulk spectrum; they must all be
irrelevant or marginal in the renormalization-group sense. It is not surprising that there exist
many possible terms that are irrelevant, as a gapless system always has infinitely many
irrelevant operators. But the fact that {\em none} of the possible terms are relevant,
and the general nature of the gapless systems, remain topics for future investigation.

\acknowledgments

We thank E. Rezayi, N. Regnault, J. Dubail, and E. Ardonne for helpful discussions, and especially
thank M. Jimbo for showing us a proof (different from ours) of Theorem 2 in the case of the
$M(3,p)$ series,
as in FJM. This work was supported by NSF grants nos.\ DMR-0706195 (TSJ and NR)
and DMR-1005895 (NR), and by EP-SRC grants nos.\ EP/I032487/1 and EP/I031014/1 (SHS).

\begin{appendix}
\section{$c=0$ case}
\label{app:c0}

In this Appendix, we give the detailed construction of the zero-energy wavefunctions for the special case
of the three-body Hamiltonian in which $c=0$, which was postponed from Section \ref{scft}; we consider
the plane only. We pointed out there that the allowed behavior for a zero-energy function is that it
must vanish like the discriminant $D_3$ in any three variables, or faster, as they come together. That
is the behavior of the Laughlin $\nu=1/2$ ground and edge states. However, there are also other functions
that are non-vanishing when two coordinates coincide, but vanish at least as fast as degree seven when
three coordinates come together.

The analysis follows the same pattern as that in Section \ref{scft}, but if anything is a little simpler.
We use the same maps $C_m$ that set the last $2m$ coordinates equal in pairs, and the same filtration
by spaces of functions $F_m$. Due to the properties just stated, the image of $F_{m+1}$ under $C_m$
($m=0$, $1$, \ldots, $\lfloor N/2\rfloor$) in this case consists of functions of the form
\bea
\prod_{i<j\leq N-2m}(z_i-z_j)^2\cdot\prod_{i\leq N-2m}\prod_{k\leq m}(z_i-Z_k)^7\cdot&&
\nonumber\\
{}\times\prod_{1\leq k<l\leq m}(Z_k-Z_l)^{14}\cdot
\Lambda_{N-2m}\otimes \Lambda_m.&&
\eea
We can construct in explicit form a set of functions that map onto these. We define
\be
\chi_3(z_1,z_2;z_3)=(z_1-z_3)^4(z_2-z_3)^3
\ee
and
\be
\chi(z_1,z_2,z_3,z_4)=(z_1-z_3)^4(z_2-z_3)^3(z_2-z_4)^4(z_1-z_4)^3,
\ee
which resemble what was used in the Gaffnian functions. Now we adopt the notation used in the functions
in Eq.\ (\ref{scft_fns}), but with $\alpha=1$, $2$, only, and $m_1=m$, $m_2=N-2m_1$, $r_1=2$, $r_2=1$.
The functions are
\bea
&&{\cal S}_z\left\{\prod_{(1,i)>(1,i')}\chi(z_{i1}^{(1)},z_{i2}^{(1)};
z_{i'1}^{(1)},z_{i'2}^{(1)})\cdot\right.\\
&&{}\times\prod_{(1,i)
>(2,i')}\chi_3(z_{i1}^{(1)},z_{i2}^{(1)}
; z_{i'1}^{(2)})\cdot\non\\
&&{}\left.\times D_{m_2}(\{z_{i1}^{(2)}\})\cdot
\prod_{\alpha=1,2}\prod_{l=1}^{m_\alpha} e_l\left(\left\{\sum_{j=1}^{r_\alpha}z_{ij}^{(\alpha)}\right\}
\right)^{n_l^{(\alpha)}}\right\}
\non
\eea
It is easily checked that these have all the necessary properties, in particular there is a
unique translation-invariant symmetric polynomial of degree seven in three variables, which is
produced by symmetrizing $\chi_3$. Hence for $n_l^{(\alpha)}$
($l=1$, \ldots, $m_\alpha$; $\alpha=1$, $2$) ranging over non-negative integers they form a
complete and linearly independent set. For $m_1=0$, we obtain the Laughlin states.

The degrees of the functions when all $n_l^{(\alpha)}$ are zero are
\bea
\lefteqn{7m_1(m_1-1)+7m_1m_2+m_2(m_2-1)}&&\nonumber\\
&=&\frac{7}{4}N(N-2) -\frac{3}{4}m_2\left(m_2-\frac{10}{3}\right).
\eea
The negative sign in the middle of the last line implies the lowest degree is found for $m_2=N$,
not for $m_2=0$ as
we found in most cases. The corresponding state is $D_N$, the Laughlin $\nu=1/2$ state. The $m_2=0$
paired wavefunction given by the above with all $n_l^{(\alpha)}=0$ has filling factor $\nu=2/7$.
The character for the space of zero-energy functions in $N$ variables follows immediately:
\be
{\rm ch}_q\,\widetilde{I}_N=\sum_{m_1,m_2:2m_1+m_2=N}\frac{q^{\frac{7}{4}N(N-2)
-\frac{3}{4}m_2(m_2-\frac{10}{3})}}
{(q)_{m_1}(q)_{m_2}}.
\ee

The pathology of this case is connected with the fact that the superconformal current $G(z)$ acting on the
vacuum produces a singular vector when $c=0$. If we wish to set all singular vectors to zero, the field
$G$ is no longer available as $\psi$ in the MR construction. It is not clear if the functions found here
can be interpreted as arising from a CFT using a MR construction. On the other hand, the Hamiltonian
is presumably gapless in the bulk in the thermodynamic limit, and we can see that
it sits on a phase boundary adjacent to a Laughlin $\nu=1/2$ phase, as adding to the special Hamiltonian
a two-body term that forbids two coordinates to be equal produces the Laughlin zero-energy
states only. In this way the system is similar to the Haffnian model \cite{green}, which is also adjacent
to the $\nu=1/2$ Laughlin state, however the special Hamiltonian and zero-energy states here are not those
of the Haffnian.

We can also give a similar treatment for the space of symmetric polynomials that are required to vanish
as degree seven or faster as any three coordinates come to the same value, simply by replacing
$D_{m_2}\Lambda_{m_2}$ for the unpaired particles in the above by $D_{m_2}\widetilde{I}_{m_2}^{\rm MR}$.
The ground state function is $D_N$ times the MR $\nu=1$ state, as found in Ref.\ \cite{src}.

\section{Proof of Theorem 1}
\label{app:lim}

First, we prove that $a(z)$ and $a(z)^\ast$ generate the VOA. We take as given these operators at
any locations $z$, and use translation invariance of ope's; we must show that the current and
Virasoro algebras are generated by these operators. To begin, we take the ope $a(z')a(z)$ as $z'\to z$
($z$ fixed),
and extract the leading term, which contains the operator $e^{2i\varphi(z)/\sqrt{\nu}}$ times
the leading operator in the $\psi\psi$ ope; the whole operator is located at any $z$. Then we fuse
this with another $a$, and repeat the process. By definition of the properties of $\psi$, after taking
the product of $k$ $a$'s, we obtain the operator $e^{ik\varphi(z)/\sqrt{\nu}}$ (times the identity in
the $\psi$ theory). Doing the same with $a^\ast$'s produces $e^{-ik\varphi(z)/\sqrt{\nu}}$. The ope of
these two operators in the charge sector (the $\varphi$ CFT) contains the current operator $j(z)$, and
the ope of two $j$s produces the stress tensor $T_\varphi(z)$. It follows that all descendants generated
by currents (``current excitations'') of the leading operator in the ope $a(z')a(z)$ as $z'\to z$ are also
in the algebra, and similarly for the leading operator when another $a(z'')$ approaches this, and so on.
(Likewise, all current descendants of $a(z)$ are also included.)

Next, we consider again the ope of $a$ with itself. In addition to the leading term referred to above,
it also contains subleading terms, each of which is a sum of the leading and sub-leading terms in
the $\psi\psi$ ope, possibly with current excitations, with weights that sum to the correct value
corresponding to the overall power of $z$. The subleading terms consisting of current excitations only
can be subtracted off, as they are already in the algebra. At each level (extra factor of $z$), that
leaves an operator containing descendants in the algebra generated by $\psi$, times current excitations.
At the first level, there is just one term, the first subleading part of the $\psi\psi$ ope (if
it is non-zero), times $e^{2i\varphi(0)/\sqrt{\nu}}$, which is therefore also in the algebra.
At the next level, the current descendants of this can be subtracted off as well, possibly leaving another
operator with only descendants in the $\psi$ algebra. Proceeding in this way, we find that all
descendants in the $\psi(z')\psi(z)$ ope, times $e^{2i\varphi(0)/\sqrt{\nu}}$, times current
excitations, are in the algebra. This can be repeated also in the ope of these
operators with a further $a(z')$, and so on. Then we obtain recursively all of the VOA $V$, defined
in Eq.\ (\ref{V_struc}). In particular, the stress tensor $T_\psi(z)$ is eventually generated.

We can now prove Theorem 1. We observe that the iterated ope of $k-1$ $a$'s, taking the leading term in
each ope, produces $e^{i(k-1)\varphi(z)\sqrt{\nu}}\psi^\ast(z)$, by definition because a further ope
with $a$ gives $e^{ik\varphi(z)/\sqrt{\nu}}$. The ope with $e^{-ik\varphi(z)/\sqrt{\nu}}$ gives $a^\ast$
as the leading term. Hence the VOA $V$ is generated by $a(z)$ and $e^{-ik\varphi(z)/\sqrt{\nu}}$, which is
the second statement in Theorem 1.

To obtain the first statement, we consider the operators $e^{iN\varphi(z)/\sqrt{\nu}}$, for
$N$ divisible by $k$, which (for $z=0$) correspond to the states $|0_N\rangle$. Acting on one of these
with $a$'s
gives the space $V^{N}$, by definition. An ope with $e^{-ik\varphi(z)/\sqrt{\nu}}$ lowers $N$ by $k$.
Hence the union of the spaces of states (or corresponding operators) $V^{-N}$ over all non-negative
multiples of $k$, or the limit $\lim_{N\to\infty}V^{-N}$ using the embeddings, gives all of the VOA $V$.
That is the first statement in Theorem 1.

\section{Proof of Theorem 2}
\label{app:thm2}

The proof is carried out in terms of the spaces of polynomials $\widetilde{I}_N$. If there are any
elements in $\cal I$ that are not generated by ${\cal I}_3$ (over $\cal A$), then this means
that the spaces of polynomials $I_N$ (the amplitudes) are determined by additional vanishing
conditions. That is, if the additional element in $\cal I$ contains $N_A$ $\xi_{-n}$s, then it determines
a map from $\widetilde{I}_N$ into $\Lambda_{N-N_A}$, the symmetric polynomials in $N-N_A$ variables.
Because of translation invariance of $\cal I$, these maps
can be organized into sets, each of which can be parametrized in terms of a center of mass variable
$Z$ for the $N_A$ variables that are removed (compare the discussion of the maps $C_m$ in
Section \ref{scft}). The ``allowed'' functions in $\widetilde{I}_N$, which are in $I_N$, are those
annihilated by these maps, in other words they lie in the kernel of all such maps (as well as in those
of the maps describing conditions obeyed by functions in $\widetilde{I}_N$). The question of whether
the additional elements (vanishing conditions) make a difference, so that $\widetilde{I}_N\subsetneq
I_N$ (a proper subset), is then the question of whether what we will call the ``pre-image'' of the maps
is non-zero.
The pre-image of a map $V_1\to V_2$ is defined as the quotient space of $V_1$ by the kernel of the map
(it is mapped surjectively onto the image in $V_2$).
For the spaces of polynomials, we can introduce a positive-definite inner product, such as that which
arises in the quantum-mechanical interpretation as functions in the LLL on the plane. Then we may choose
representatives for a basis of the pre-image as vectors orthogonal to the kernel, with respect to the
inner product. If we use second-quantization methods, with $\psi(z)$ as the destruction operator for
bosons in the LLL, then we can conveniently write the maps in the form
\bea
\lefteqn{\kappa_{N_A,f}(Z)=}&&\\
&&\int_{\sum_i z_i=N_AZ}\prod_i(\mu(z_i)d^2z_i) \overline{f(z_1,\ldots,z_{N_A})}
\psi(z_1)\cdots\psi(z_{N_A}),\nonumber
\eea
where $f(z_1,\ldots,z_{N_A})$ is a homogeneous translation-invariant symmetric polynomial (that is,
it is annihilated by $\ell_{-1}$) in the orthogonal complement
of the kernel, and $\mu(z)$ is the integration measure for the geometry in use (e.g.\
$\mu(z)=e^{-|z|^2/2}$ for the plane). The domain is restricted so that the center of mass variable $Z$
has any fixed complex value. If desired, this $Z$-dependent operator can be expanded in powers of $Z$
to obtain a set of distinct operators $\kappa_{N_A,f,m}$, $m=0$, $1$, $2$, \ldots. Notice
that $\kappa_{N_A,f}(Z)$ written in this form has no direct dependence on $N$.

We can first show that the ground state (lowest-degree) function in $\widetilde{I}_N$ lies in
the kernel of these additional maps $\kappa$ for any $N$. The lowest degree function is unique, so if
it is not in the kernel of one of them, the lowest degree in $I_N$ is higher than that in
$\widetilde{I}_N$, which is not the case. (Indeed, for the ${\cal N}=1$ SCFT functions, we know
that the ground state function
in $\widetilde{I}_N$ is in $I_N$ by other means \cite{simon09}; this in fact motivated the conditions
expressed as ${\cal I}_3$.) The idea for the remainder of the proof is that any functions in
$\widetilde{I}_N$ not in the kernel of any additional maps involved must lie in the subspaces
with $2m_2+m_3>0$, and in fact have the same form for all $N$.

To demonstrate this, we first construct maps $p_{N,m}:\widetilde{I}_N\to \Lambda_{N-2m}$. It is clear
that the image of each one is in $\widetilde{I}_{N-2m}$, and we will argue that the image space is
isomorphic to a copy of $\widetilde{I}_{N-2m}$.

$p_{N,m}:\widetilde{I}_N\to\widetilde{I}_{N-2m}$ can be defined in either of two ways,
which are equivalent up to a constant factor. One way is let the first $2m$ variables in the function
tend to zero, one after another, and at each step project onto the power that is the leading one for
that step in the case of the ground state wavefunction (the powers can thus be read off from the
known function). The other way is to project onto the ground state wavefunction in $\widetilde{I}_{2m}$
for the first $2m$ variables. The projections can be defined using the quantum-mechanical inner product,
and second-quantization, similar to $\kappa_{N,f,0}$ above, but using (in the second version) the ground
state function in place of $f$, and with $2m$ destruction operators. In this approach, we see at once
that the two linear maps $p_{N,m}$ and $\kappa_{N_A,f}$ commute.

From the wavefunctions in Eq.\ (\ref{scft_fns}), we can see that for a function in the $m_1$, $m_2$,
$m_3$ space there is a non-zero contribution to the image under $p_{N,m}$ only if $m\leq m_1$,
and to obtain this image, all of the first $2m$ variables must be assigned to type $\alpha=1$ under
the sum over permutations that defines the symmetrization. Without loss of generality (that is,
by relabeling permutations in terms in the image), we can think of the variables removed as lying
in the first $m$ rows of the partition. From the definition of $p_{N,m}$, we see that the functions
in the image are of the form $\prod_i z_i^{6m} g$, where $i$ runs over (and $g$ is a function of)
the remaining $N-2m$ variables, and $g$ lies in $\widetilde{I}_{N-2m}$ by inspection of the
functions ($\prod_i z_i^{6m} g$ itself lies in $\widetilde{I}_{N-2m}$, because the latter is a
module over the symmetric polynomials). Moreover, $p_{N,m}$ is surjective onto this space
$\prod_i z_i^{6m} \widetilde{I}_{N-2m}$. In low degrees, it is also injective. To be injective, the
degree of the initial function must be low enough so that
no elementary symmetric polynomials $e_l$ in the type $\alpha=1$ variables
of degree $l> N-2m -2m_2-m_3$ (in the parametrization of subspaces as in Section \ref{scft}) can be
used. (The images of those generators are not algebraically independent of the lower degree ones
in the image space.) We remark that $p_{N,m}=p_{N,1}\circ\cdots\circ p_{N,1}$, the composite of
$m$ maps. This completes the construction and properties of $p_{N,m}$. Also, we
may view multiplication by $\prod_i z_i^{6m}$ on $\widetilde{I}_{N-2m}$
more formally as an injective map $r_{N,m}$ mapping $\widetilde{I}_{N-2m}\to\widetilde{I}_{N-2m}$,
with image $\prod_i z_i^{6m} \widetilde{I}_{N-2m}$.


Now we conclude the proof of Theorem 2. We have to show that there are functions
in $\widetilde{I}_N$
that are not annihilated by some of the operators $\kappa$, and that these occur in low degree relative
to the ground state for all sufficiently large $N$. In that case, it will follow that the dimensions
of the spaces appearing as coefficients in the (shifted) character in the limit are less than
in the character of ${\rm ch}_q V$, which is a contradiction.

It will be sufficient to show that there is some basis function in $\widetilde{I}_N$ not annihilated by
some $\kappa$, for some $N$. By hypothesis there
is some $N_A$ for which there is such a basis vector in $\widetilde{I}_{N_A}$. We will use the
(non-orthonormal) basis set of functions in Eq.\ (\ref{scft_fns}) for $N=N_A$. Suppose that
$g_{N_A,m_2,m_3,\{n_l^{(1)}\},\{n_l^{(2)}\},\{n_l^{(3)}\}}$ (in obvious notation) is such a function,
not annihilated by $\kappa_{N_A,f}(Z)$; we will denote it by $g$ for brevity. We can choose $N_A$
and the degree of $g$ to be the minimal values for which this holds; in particular, we can assume
that $g$ is annihilated by $\ell_{-1}$. In that case, (i) it is $\kappa_{N_A,f,0}$ that does not
annihilate $g$; (ii) $g$ has the same degree as $f$; and (iii) $\kappa_{N_A,f,m'}g=0$ for all $m'>0$.

We will show there is a corresponding function in $\widetilde{I}_N$ for $N=N_A+2m$, for any positive $m$,
that is not annihilated by $\kappa_{N_A,f}(Z)$. First, we can use $r_{N,m}$ to map
$g$ into $\prod_i z_i^{6m} \widetilde{I}_{N-2m}$.
In the product $\prod_i z_i^{6m}$, each $z_i$ can be expanded as $z_i=Z+(z_i-Z)$. If we choose the term
with all $z_i$s replaced by $Z$, the factors of $Z$ simply raise the center of mass angular momentum,
and $Z^{6m(N-2m)}g$ is not annihilated by $\kappa_{N,f,6m(N-2m)}$. In the remaining terms, the degree
in the internal (relative) variables is raised, so they are annihilated by $\kappa_{N_A,f,m'}$ for
all $m'$. Thus $\kappa_{N,f,6m(N-2m)}\prod_i z_i^{6m}g\neq0$.

$p_{N,m}$ is surjective onto $\prod_i z_i^{6m} \widetilde{I}_{N-2m}$, so there is a corresponding
function $\hat{g}$ in $\widetilde{I}_N$ that maps to $\prod_i z_i^{6m}g$, that is $p_{N,m}\hat{g}=\prod_i
z_i^{6m}g$. $\hat{g}$ can be taken to be the basis function with labels
$m_2$, $m_3$, $\{n_l^{(1)}\}$, $\{n_l^{(2)}\}$, $\{n_l^{(3)}\}$. This function $\hat{g}\in \widetilde{I}_N$
is not annihilated by $\kappa_{N,f,6m(N-2m)}$, because if $\kappa_{N,f,6m(N-2m)}\hat{g}=0$, then
$p_{N,m}\kappa_{N,f,6m(N-2m)}\hat{g}=\kappa_{N,f,6m(N-2m)}p_{N,m}\hat{g}=0$, which contradicts what was
just shown. We notice that the difference of the degree of $\hat{g}$ from the degree of the $N$-particle
ground state is determined by the numbers $m_2$, $m_3$, $\{n_l^{(1)}\}$, $\{n_l^{(2)}\}$,
$\{n_l^{(3)}\}$, and is independent of $N$.

We have shown that any additional relations in $\cal I$, or vanishing conditions needed to define $I_N$,
other than those generated by ${\cal I}_3$, cause some functions to disappear at fixed degree relative to
the ground state for all $N=N_A+2m$, and so the character in the limit becomes smaller. But this
contradicts
the agreement that was established. Hence we conclude that $\cal I$ is generated by ${\cal I}_3$, and
the vanishing conditions in three variables which define $\widetilde{I}_N$ also define $I_N$, and
so these spaces are the same, which is Theorem 2. We remark that translation invariance played
an essential role in the proof. Similar proofs can be constructed in other examples, with the number
$2$ replaced by $k$ for the order $k$ examples, a suitable change in the power $z_i^{6m}$, and so on
{\it mutatis mutandis}.

\end{appendix}

\end{document}